\documentclass{article}
\usepackage{amsmath,amssymb,amsfonts,mathtools}
\usepackage{graphicx}
\usepackage{textcomp}
\def\BibTeX{{\rm B\kern-.05em{\sc i\kern-.025em b}\kern-.08em
    T\kern-.1667em\lower.7ex\hbox{E}\kern-.125emX}}

\usepackage{physics}
\usepackage{bm} % bold greek letters
\usepackage{pdfpages}
\usepackage[inline]{enumitem}
\usepackage{subcaption}
\usepackage{booktabs}

\newcommand{\R}[0]{\mathbb{R}}
\newcommand{\C}[0]{\mathbb{C}}
\newcommand{\Hilbert}[0]{\mathcal{H}}

\newcommand{\X}[0]{\mathcal{X}}

\newcommand{\Y}[0]{\mathcal{Y}}
\newcommand{\D}[0]{\mathcal{D}}

\newcommand{\vectheta}[0]{\mathbf{\bm{\theta}}}
\newcommand{\vecx}[0]{\mathbf{\bm{x}}}
\newcommand{\vecw}[0]{\mathbf{\bm{w}}}
\renewcommand{\emph}[1]{\textit{#1}} % IEEE don't show italics

\usepackage{authblk}
    
\begin{document}

\title{The Quantum Path Kernel: a Generalized Quantum Neural Tangent Kernel for Deep Quantum Machine Learning}

\author[1]{Massimiliano~Incudini}
\author[2]{Michele~Grossi}
\author[3]{Antonio~Mandarino}
\author[2]{Sofia~Vallecorsa}
\author[1]{Alessandra~Di~Pierro}
\author[4]{David~Windridge}

\affil[1]{Department of Computer Science, University of Verona, Verona 37134, Italy}
\affil[2]{European Organization for Nuclear Research (CERN), Geneva 1211, Switzerland}
\affil[3]{International Centre for Theory of Quantum Technologies (ICTQT), University of Gdansk, 80-309 Gdańsk, Poland}
\affil[4]{Department of Computer Science, Middlesex University, The Burroughs, London, NW4 4BT, UK}

\date{}

%\titlepgskip=-15pt

\maketitle

\begin{abstract}
Building a quantum analog of classical deep neural networks represents a fundamental challenge in quantum computing. 
A key issue is how to address the inherent non-linearity of classical deep learning, a problem in the quantum domain due to the fact that the composition of an arbitrary number of quantum gates, consisting of a series of sequential unitary transformations, is intrinsically linear. %irrespective of circuit complexity. 
This problem has been variously approached in the literature, principally via the introduction of measurements between layers of unitary transformations.
In this paper, we introduce the Quantum Path Kernel, a formulation of quantum machine learning capable of replicating those aspects of deep machine learning typically associated with superior generalization performance in the classical domain, specifically, \emph{hierarchical feature learning}. Our approach generalizes the notion of Quantum Neural Tangent Kernel, which has been used to study the dynamics of classical and quantum machine learning models. 
The Quantum Path Kernel exploits the parameter trajectory, i.e. the curve delineated by model parameters as they evolve during training, enabling the representation of differential layer-wise convergence behaviors, or the formation of hierarchical parametric dependencies, in terms of their manifestation in the gradient space of the predictor function. 
We evaluate our approach with respect to variants of the
classification of Gaussian XOR mixtures - an artificial but emblematic problem that
intrinsically requires multilevel learning in order to achieve optimal class separation. 
\end{abstract}

\section{Introduction}

Bridging classical deep neural networks %\cite{lecun2015deep} 
and quantum computing represents a key research challenge in the field of \emph{quantum machine learning} \cite{wittek2014quantum,schuld2015introduction}. 
The potential for improvement offered by  quantum computing in the machine learning domain may be characterized in terms of its impact on algorithmic efficiency, generalization error, or else its capacity for treating quantum data \cite{huang2021power}.

A notable recent result in the field has been the introduction of the concept of \emph{variational quantum algorithms} and the related  neural network analog referred to as the \emph{quantum neural network} (QNN) \cite{abbas2021power}.
This, in essence, consists of a feature map encoding data into a quantum Hilbert space upon which certain parameterized unitary rotations are applied  prior to final measurement in order to obtain a classification or regression output. The system as a whole is then optimized by classical methods. 
Such models provably lead to a computational advantage over classical models on certain artificial tasks \cite{liu2021rigorous},  and in respect to the analysis of specific physical systems \cite{huang2022quantum}. It has been quantitatively shown that QNNs can be trained faster than their classical analogues \cite{abbas2021power}.
However, QNNs remain problematic in various respects. 
One limitation arises from the so-called barren plateau problem  \cite{mcclean2018barren}, in which the variance of the gradient vanishes exponentially  with the system size as the parameterized transformation becomes increasingly expressive \cite{holmes2021connecting}. A number of approaches, including layer-wise training of quantum neural networks \cite{skolik2021layerwise}, have been proposed to mitigate the issue.

A second problematic aspect of QNNs, and the one that constitutes our principal focus here, is the linearity of the dynamics of quantum systems. Concatenations of linear unitary transformations remain unitary and thus  `stacked' quantum transformations, in effect, collapse to a single linear transformation, appearing to rule out de facto the hierarchical feature learning of classical deep neural networks, which relies on non-linearities to separate feature layers. This property makes the QNN essentially a kernel machine \cite{schuld2021supervised}. 
In terms of the predictor function, however, the QNN is composed of multiplications of rotation operators parameterized by both the feature and model weights. The nonlinearity of projections of rotation operators can  be exploited to replicate a very constrained form of non-linearity for feature learning \cite{liu2022representation}. Another strategy is to introduce nonlinearity via the measurement operation, i.e. a \emph{dissipative QNN}  \cite{sharma2020trainability}. Both approaches involve the projection the quantum state into a subspace of the original Hilbert space.

Much of the recent study of the dynamics of deep neural networks in the classical realm has focused on the Neural Tangent Kernel (NTK) \cite{jacot2018neural}
which represents the network in terms of the corresponding training gradients in the model parameter space.
The NTK hence approximates the behavior of predictors via a linear model. It is often therefore applied to study neural networks in their asymptotic, infinite-width, limit. In this regime, the network exhibits  \emph{lazy training} \cite{chizat2019lazy}, i.e. parameter gradients remain at their initial values during the entirety of training. The NTK thus accurately characterizes the dynamics of such infinite-width neural networks, but is otherwise only an approximation \cite{liu2020linearity}. The difference in test error between the predictor and its linearized version depends on the problem structure \cite{ghorbani2021neural}, with hierarchical feature learning capability being crucial to obtaining superior performance \cite{chen2020towards}. However, the kernel nature of the NTK means that it shares with quantum computing a ready interpretation within a Hilbert space, and is thus of considerable interest within quantum machine learning. The first explicit application of NTK to quantum neural networks, the \emph{quantum neural tangent kernel} (QNTK) was given in \cite{shirai2021quantum}. 

In this paper, we propose a method for overcoming the de facto lack of hierarchical feature learning capability in QNNs. 
We propose the application of Path Kernels \cite{domingos2020every} to QNNs, which we call the \emph{Quantum Path Kernel} (QPK). Such an approach generalizes the QNTK so that the resulting kernel is representative of the ensemble of NTKs calculated over the full parameter path trajectory, i.e. the function describing the evolution of model parameters over time, including implicitly any parametric evolutions corresponding to hierarchical feature learning. 
We show experimentally an increased expressivity of the resulting model relative to linearized equivalents, evaluating our method on the Gaussian XOR mixture classification problem. For this problem, finite-width neural networks have both theoretically and empirically shown to be close-to-optimal performance whereas linear NTK models fail \cite{refinetti2021classifying}, suggesting that it cannot be effectively resolved without implicating multilevel learning behavior. Furthermore, we discuss possible improvements for the proposed approach, which can be obtained by considering only the contribution of the parameter gradient path that gives rise to the most decorrelated feature representation. These specifically corresponds to the contributions associated with the maximally nonlinear point of the parameter path, corresponding to the largest (positive or negative) eigenvalues of the Hessian of the predictor function \cite{ghorbani2019investigation}. We further enhance the decorrelation between feature representations via a stochastic, noisy, or non-gradient-descent-based training algorithm in which the averaging  operation between decorrelated representations allows us to interpret the model as an ensemble technique.

The paper is structured as follows. 
In Section~\ref{sec:background} we briefly review the necessary conceptual background. 
In Section~\ref{sec:quantumpathkernel} we present the Quantum Path Kernel and discuss the hierarchical feature learning of the induced model. 
In Section~\ref{sec:experiments} we demonstrate how this leads to superior performance in solving the Gaussian XOR mixture classification problem. 
In Section~\ref{sec:conclusion} we draw our conclusions and present directions for further work. 

\subsection{Contributions} 
\begin{itemize}
    \item We propose the \emph{Quantum Path Kernel} as a mechanism for building hybrid classical/quantum machine learning models which are able to emulate the hierarchical feature learning structure of deep neural networks without violating the underlying linearity of the quantum dynamics.
    \item We provide numerical evidence of the superior performance of the Quantum Path Kernel compared to the QNTK on the Gaussian XOR Mixture problem, which is Bayes optimally soluble only through implicating layerwise nonlinear separability.
    \item We consider the importance of the extraction of non-correlated feature representations corresponding to maximally varying portions of the parameter gradient path. % We suggest  stochastically perturbing gradient descent  as a mechanism for inducing further decorrelation between feature representations.
\end{itemize}

\subsection{Related works} The introduction of the NTK by \cite{jacot2018neural} has marked a significant step in the theory of machine learning, sheding new light on discussions regarding the relative  performance of linear and nonlinear models. For example, \cite{ghorbani2021neural} suggests that tasks in which kernel methods (including NTK) perform worse than neural networks are those in which the kernel suffers from the curse of dimensionality whereas neural networks, in learning some useful lower dimensional representation, do not. One example of such a problem is the Gaussian XOR Mixture classification task \cite{refinetti2021classifying}. Furthermore, linearized models have been shown to perform slightly worse than wide (i.e. large, but non-infinite) neural networks on CIFAR-10 benchmark \cite{arora2019exact}, with the gap between the approaches increasing for finite width networks \cite{bai2020taylorized}. 

In relation to  quantum computation, researchers have spent substantial effort on the limitations imposed by the linear dynamics of quantum systems. Authors in \cite{schuld2014quest}  review  early approaches to the formulation of nonlinear quantum machine learning models: some have focused on developing a \emph{quantum perceptron} equivalent or \emph{quantum neuron}, i.e. a candidate building block for the quantum analogue of neural networks; \cite{schuld2015simulating} uses phase estimation to implement the functioning of a step function; \cite{cao2017quantum, hu2018towards} propose to exploit the RUS (repeat until success) policy to mimic the behaviour of tangent and sigmoid activation functions, while \cite{gili2022introducing} uses RUS to construct a Born machine; \cite{tacchino2019artificial} emulates the nonlinearity of perceptrons using measurements. In relation to QNNs, \cite{sharma2022trainability} propose dissipative QNNs in which the  nonlinearity is obtained via intertwining measurements between unitary gates;  \cite{guo2021nonlinear, holmes2021nonlinear} propose the use of a larger Hilbert space to implement the nonlinear transformation, while \cite{daskin2018simple} exploits the exponential form of unitary gate to achieve periodic activation functions. Finally, non-linear models of quantum mechanics have been conjectured by \cite{Weinberg1989Precision}, although these violate some computational complexity assumptions \cite{abrams1998nonlinear}. 

\section{Background}\label{sec:background}

This section briefly introduces the key concepts and notations in relation to Deep Learning and Quantum Machine Learning through which we  develop our results.
We denote by $\D = \{ (\vecx_i, y_i) \}_{i=1}^n \subseteq \X \times \Y$ a labelled dataset of pairs that are i.i.d. sampled from an unknown probability distribution. We indicate the data vector space with $\X = \R^d$, and the target space with either $\Y = \R$ or $\Y \subseteq \mathbb{Z}, |\Y| < \infty$ for regression or classification tasks, respectively. We indicate uniform sampling from a uniform discrete distribution with $\sim \{ v_i \}_{i=1}^n$ and sampling from a normal distribution of mean $\mu$ and variance $\sigma^2$ with $\sim \mathcal{N}(\mu, \sigma)$.

\subsection{A primer on quantum machine learning models} \label{sec:background:qm}

Here we fix the notation for our quantum machine learning models. The state of a quantum system of $m$-qubits is described by a density matrix $\rho \in \Hilbert \equiv \C^{2^m \times 2^m}$. The initial state of a quantum computation is denoted by $\rho_0 = \ketbra{0}{0}$, and the (possibly parametric) unitary transformations by $U, V, W$. Any parametric unitary can be written as
\begin{equation}
    U(\vectheta) = \exp{-i 
    \sum_{k=1}^m f_j(\vectheta) \sigma_{\alpha_1, ..., \alpha_k}^{(q_1, ..., q_k)}
    },
\end{equation}
where $\alpha_i \in \{\textsc{x}, \textsc{y}, \textsc{z}, \mathbf{1}\}$ for $i=1,\ldots,k$, and $\sigma_{\alpha_1, ..., \alpha_k}$ is a tensor product of one or more corresponding Pauli matrices applied to qubits $q_1, ..., q_k$. The same transformation may be interpreted as a rotation and be equivalently denoted by $R_{\alpha_1, ..., \alpha_k}^{(i_1, ..., i_k)}(\vectheta)$, where $\vectheta \in \R^P$ are rotational angles. A \emph{quantum neural network} is a function of the form\footnote{The most general form of QNN proposed is the \emph{data re-uploading} QNN, which allows the interspersing of data encoding and trainable transformations. Such a form, however, does not add any computational power to the standard QNN approach \cite{jerbi2021quantum}.}:
\begin{equation}
    f(\vecx; \vectheta)
    = \Trace[\rho_{\vecx, \vectheta} O]
    = \Trace[V^\dagger(\vectheta) U_\phi^\dagger(\vecx)\rho_0 U_\phi(\vecx) V(\vectheta) O], 
    \label{eq:ymodel}
\end{equation}
where $O$ indicates any measurement operator.
Both the matrices $U$ and $V$ are decomposed in single and two-qubits parametric rotations interspersed with non-parametric gates (e.g. CNOT).

\subsection{Notions of nonlinearity in classical and quantum learning models}\label{sec:background:linearity}

With respect to both kernel machines and layerwise deep learning, the concepts of \emph{linear model}, \emph{nonlinear model}, and \emph{feature learning} that we utilize here are as  formalized in \cite{roberts2021principles}. A \emph{linear model} is thus a function of the form:
\begin{equation}\label{eq:linmodel}
    f(\vecx; \vectheta) = \sum_{j = 1}^p \theta_j \phi_j(\vecx),
\end{equation}
where $\{ \phi_j : \X \to \R \}_{j=0}^p$ are the \emph{feature functions}, whose values corresponds with the model features. We might consider an additional feature $\phi_0 \equiv 1$ that incorporates the bias. The formula in Equation \ref{eq:linmodel} is linear with respect to the space of the parameters\footnote{This formalism allows us to exploit even infinite dimensional Hilbert spaces, such as the one implemented by the Gaussian feature map or RBF, mapping $\vecx$ to a multivariate Gaussian of mean $\vecx$ and fixed covariance, existing in the space of square-integrable multivariate functions.} $\Hilbert \equiv \R^p$; in fact, we can interpret the function as an inner product in that space, i.e.
\begin{equation}
    f(\vecx; \vectheta) = \langle \vectheta, \bm{\phi}(\vecx) \rangle_{\R^p},
\end{equation}
with $\vectheta = (\theta_1, ..., \theta_p)$ and $\bm{\phi}(\vecx) = (\phi_1(\vecx), ..., \phi_p(\vecx))$. 
The optimal parameters of such a model can be found analytically by solving the linear regression problem over the Mean Squared Error loss, which is a convex, quadratic function of the parameters. The \emph{representer theorem} guarantees that the optimal solution is a span of the $m$ data points of the training set, which is independent from the dimensionality $n$ of the space $\Hilbert$. Obviously, a model which is linear in the parameters may well behave nonlinearly with respect to the original feature space $\X$, due to the feature functions.

A \emph{nonlinear model} is a function of the form:
\begin{equation}\label{eq:nonlinmodel}
    f(\vecx; \vectheta) = \sum_{j = 1}^p \theta_j \phi_j(\vecx)
    + \frac{\epsilon}{2} \sum_{j,k=1}^p \theta_{j} \theta_k \psi_{j,k}(\vecx) 
    + \frac{\epsilon}{3!} \sum_{j,k,\ell=1}^p \theta_{j} \theta_k \theta_\ell \psi_{j,k,\ell}(\vecx) 
    + \cdots 
\end{equation}
The higher-order terms of the expansion are characterized by their own set of features, e.g. $\{ \psi_{j,k} : \X \to \R \}_{j, k = 1}^p$ for the second order term. The elements of such sets are unique up to a permutation of their variables, thus the terms $1/2!, 1/3!, ...$ compensate the multiple counting of such elements in Equation \ref{eq:nonlinmodel}. The term $\epsilon \ll 1$ adjusts the contribution of the nonlinear terms. If the model is truncated to the second term it is denoted as \emph{quadratic model}. In such a case, the loss function is quartic, thus we cannot find analytically the optimal parameters as in the linear regression. The dynamic of such a model is described by
\begin{align}
    & f(\vecx, \vectheta+d\vectheta) \\
    & = f(\vecx, \vectheta) + \sum_{j=1}^p d\theta_j \left[ \phi_j(\vecx) + \epsilon \sum_{k=1}^p \theta_j \psi_{j,k}(\vecx) \right] + \frac{\epsilon}{2} \sum_{j,k=1}^p d\theta_{j} d\theta_k \psi_{j,k}(\vecx) \\
    & = f(\vecx, \vectheta) + \sum_{j=1}^p d\theta_j \phi^{E}_j(\vecx; \vectheta) + \frac{\epsilon}{2} \sum_{j,k=1}^p d\theta_{j} d\theta_k \psi_{j,k}(\vecx) 
\end{align}
where $\phi^{E}$ are \emph{effective feature functions}, i.e. features that depend on, and evolve with, the model parameters, which are learnt during the optimization phase. This behaviour can be generalized to consider terms of even higher orders: the presence of order $n$ terms make the feature functions of order $n-1$, effectively, which may further influence the lower order terms. Models having effective feature functions have \emph{feature learning} capabilities. A \emph{deep learning model} is both capable of feature learning and composed of several nonlinear modules arranged in a hierarchical fashion \cite{lecun2015deep}; such that differing layers can follow differing (albeit hierarchically conditioned) gradient paths.

Turning to QNNs, the quantum model
\begin{equation}
    f(\vecx; \vectheta) = \Trace[\rho_{\vecx, \vectheta} O], 
\end{equation}
where $\rho_{\vecx, \vectheta} = V^\dagger(\vectheta) U^\dagger_\phi(\vecx) \ketbra{0}{0} U_\phi(\vecx) V(\vectheta)$, and $O$ Hermitian observable, is a linear model in the space of density matrices of the quantum system $\Hilbert$: the trace operation $\Tr[A^\dagger B]$ is an inner product for the space of matrices $\C^{k \times k}$. 
Such a property implies that the construction of a layer-wise architecture for $v$, i.e. $v(\vectheta) = \prod_i V_i(\vectheta)$ effectively collapses to a single operation: this may add more degrees of freedom to the linear transformation\footnote{depending on the generators involved and up to a maximum of $4^n-1$ (where $n$ number of qubits)} but cannot make the model nonlinear in $\Hilbert$. 

However, in terms of the predictor function $f(\vecx; \vectheta)$, the quantum model does not necessarily fit the form set out Equation \ref{eq:linmodel} since the parameters of the QNN model, namely the angle of rotation operation (in the form of imaginary exponential function), are subject to the trace operation. Thus, for example, consider a single-qubit quantum model acting on a single input $\vecx \in \R^1$, depending on a single parameter $\vectheta \in \R^1$, with feature map  $U_\phi(\vecx) = \exp(-i x \sigma_x)$, variational form  $V(\vectheta) = \exp(-i \theta \sigma_x)$ and measurement operator i $O = \sigma_z$, in which case $f(\vecx; \vectheta)$ has the form:
\begin{equation}\label{eq:simplemodel}
    f(\vecx; \vectheta) = 
    \Trace\left[
    \begin{psmallmatrix}
    \cos^2(\theta+x) & -i\sin(\theta+x)\cos(\theta+x) \\
    \frac{1}{2}i\sin(2(\theta+x)) & \sin^2(\theta+x)
    \end{psmallmatrix}
    \begin{psmallmatrix}
        1 & 0 \\ 0 & -1
    \end{psmallmatrix}
    \right]
    = \cos(2(\theta + x))
\end{equation}
which is nonlinear in its weights. Clearly, if we were to consider a model other than a QNN then the predictor function  would change, for example as in \cite{tacchino2019artificial}, however it does not alter our argument here.

To recap, a QNN is a linear model in the Hilbert space of the density matrices due to the linearity of the evolution of closed quantum systems. However, its predictor is nonlinear in the parameter $\vectheta$ since its structure results in a composition of trigonometric functions. This potentially allows a limited degree of representational learning capability if aggregated layer-wise (limited in the sense of applying only to a highly constrained set of activation functions). However, due to the Lie algebraic equivalence of any given sequence of quantum transformations to some single unitary operation in the absence of the trace operation, we are still not able to characterise truly deep models in the quantum domain.  

\subsection{Characterization of model dynamics through the Neural Tangent Kernel}

The output $f(\vecx; \vectheta)$ of a machine learning model trained via (possibly stochastic) gradient descent can be approximated as a first-order Taylor expansion $f(\vecx; \vectheta) \approx f(\vecx; \vectheta_0) + \nabla_\vectheta f(\vecx; \vectheta_0)(\vectheta - \vectheta_0)$. Such an approximation allows the representation of  machine learners  as linear (kernel) models via the Neural Tangent Kernel (NTK, \cite{jacot2018neural}):
\begin{equation}
    k_\text{ntk}(\vecx, \vecx'; \vectheta) 
    = 
    \nabla_\vectheta f(\vecx; \vectheta) 
    \cdot 
    \nabla_\vectheta f(\vecx'; \vectheta)
\end{equation}

Such a tool has been used in \cite{chizat2019lazy} to characterize the dynamics of infinite-width neural networks, in which the NTK is independent of the random initialization and constant in time. On a coarse level of detail, we can assert that model training in \emph{lazy-training regime}, i.e. when the evolution of $\vectheta(t)$ during the training of the model $f(\vecx, \vectheta)$ closely follows the tangent path, can be decently approximated by the NTK. A more detailed analysis in \cite{liu2020linearity} has revealed that the NTK is constant if and only if the model is linear (in its parameters). Such a result allows us to quantify the nonlinearity of a model through its Hessian norm of the predictor function: if $\lVert H_f \rVert \ll \lVert \nabla_w f \rVert$ then the model is nearly linear. This has been used in \cite{liu2022representation} to analyze the behaviour of the QNNs in the lazy training regime. 

\section{The Quantum Path Kernel Framework}\label{sec:quantumpathkernel}

No extant quantum method is thus able to fully capture the deviations from gradient path linearity   manifested by empirically optimal learners in the classical domain. Hence, in order to encompass the concepts of hierarchicality and  feature learning in (implicitly kernel-based) quantum machine learning models, we here introduce for the first time in the quantum realm a key idea of Domingo's \cite{domingos2020every}, namely \emph{Path Kernelization}. 

Within this paradigm, for any machine learning model $f_{\bar\vectheta}(\vecx)$ whose parameters $\bar\vectheta$ are learned from a set $\D = \{ (\vecx_i, y_i) \}_{i=1}^n$ by gradient descent via a differentiable loss function, it is possible to express the resulting  (i.e. trained) classifier as:
\begin{equation}
    f_{\bar\vectheta}(\vecx) \approx  
    \sum_{i=1}^n \alpha_i(\vecx) \, k_\text{path}(\vecx, \vecx_i; \bar\gamma) + \alpha_0(\vecx)
\end{equation}
where
\begin{equation}
    k_\text{path}(\vecx, \vecx'; \gamma) = 
    \int_{\gamma} 
    \nabla_\theta 
    f(\vecx; \vectheta)
    \cdot 
    \nabla_\theta 
    f(\vecx'; \vectheta)
    \; d\theta
    \label{eq:pk}
\end{equation}
is the \emph{Path Kernel}, i.e. the line integral of $k_\text{ntk}$ over the multidimensional curve representing the evolution of the parameters $\vectheta = \gamma(t), t \in [0, T]$ during training, with $\bar\vectheta = \bar\gamma(T)$. In general, chain rule dependencies arising from the specifics of the architecture of the network will imply hierarchical dependencies among the parameters during learning. The result holds even for stochastic gradient descent optimization, in which case Equation~\ref{eq:pk} is a stochastic integral. However, it is not immediately clear that this path integration obeys Mercer's conditions; while it is generally true that a convex sum over Mercer kernels  is itself a Mercer kernel, the path over which we are integration is here \emph{dependent on the training objects}. We therefore dedicate Appendix~\ref{apx:path} to proving that the Path Kernel is effectively  Mercer, and set out the pseudocode for its construction. 

It is thus central to our argument to examine the parameter path $\gamma$ and its morphological evolution. For linear models, assuming a vanilla gradient descent training over a convex loss function $\mathcal{L}$, the parameter path is described by a linear vector $\{ (1 - t) \vectheta_0 + t \vectheta_f \mid t \in \R \}$ where $\vectheta_0 \in \R^p$ are the parameters at their initialization, and $\vectheta_f \in \R^p$ are the parameters at their convergence on the (ideally global) minima of $\mathcal{L}$. In such a case, it is immediately possible to check that the derivative of the linear model $\nabla_\theta f$ is independent of $\theta$, and thus that the NTK is constant. For nonlinear models, the loss function $\mathcal{L}$ may become non-convex and $\gamma$ is not constrained to be a linear trajectory. In this latter case, both the $\nabla_\theta f$ and the NTK will vary in time. 

\begin{figure*}[tbp]
    \centering
    \includegraphics[width=\textwidth]{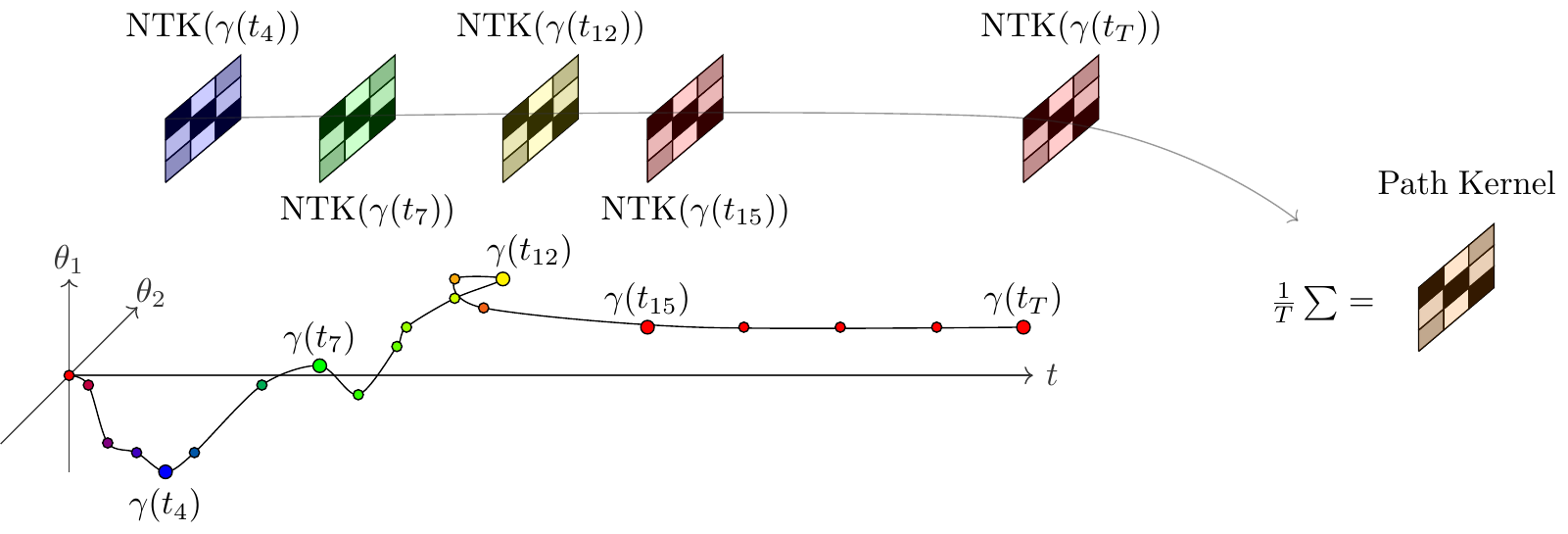}
    \caption{Computation of the Path Kernel. 
    \emph{Bottom left}: A typical parameter trajectory $\gamma$ is depicted, representing parametric evolution during the training phase. 
    \emph{Top left}: as $\vectheta$ evolves, it gives rise to differing NTK matrices, corresponding to distinct  representations of the data. Such a sequence of matrices thus give rise to a hierarchical stack of representations in the feature learning regime.
    \emph{Middle}: as the training approaches convergence, subsequent matrices become similar to each other, and thus their corresponding representations are correlated. 
    \emph{Right}: the Path Kernel constitutes the average over these representations.    
}
    \label{fig:scheme_path}
\end{figure*}

In this work, we will not focus on the possible role of Path Kernels in approximating nonlinear models.  
Instead, we shall exploit the intrinsically hierarchical structure of the Path Kernel to implement a hybrid deep machine learning model within a quantum neural network setting. We depict the construction of this object in Figure~\ref{fig:scheme_path}. The parameter trajectory for a nonlinear model is described by a complex, non-straight curve. Each point of the parameter path $\vectheta_t = \gamma(t)$ may be used to define a new kernel representation for the training data, namely $k_\text{ntk}(\vecx, \vecx'; \vectheta_t)$. We can then define a sequence of kernels stacked in a hierarchical way (whose structure, in passing, resembles the layers of a deep neural network, though this observation is peripheral to the argument being made here). Thus, each new ``layer" is a source of representation learning: the new representation (i.e. kernel matrix) is the result of an optimization process that further adapts the previous representation to the given data discrimination problem (which resembles, though is again not equivalent to, classifier boosting). 

%The structure of the Path Kernel having the hierarchical feature learning characteristics that lack a quantum neural network, 
It thus becomes possible, via explicit substitution for the corresponding Quantum NTK previously defined, to construct a Quantum Path Kernel (QPK) as follows: 
\begin{align}
    & k_\mathrm{qpk}(\vecx, \vecx'; \gamma) \nonumber \\
    &
    = 
    \frac{1}{\lVert \gamma \rVert}
    \int_{\gamma}
    \nabla_\vectheta \expval{
        V^\dagger(\vecx)
        U^\dagger(\vectheta)
        O
        U(\vectheta)
        V(\vecx)
        }{0}^T
    \cdot \nabla_\vectheta \expval{
        V^\dagger(\vecx')
        U^\dagger(\vectheta)
        O
        U(\vectheta)
        V(\vecx')
        }{0}
    \, d\vectheta \label{eq:pathker1} \\
    & = 
    \frac{1}{\lVert \gamma \rVert}
    \int_{0}^T
    \nabla_\vectheta \expval{
        V^\dagger(\vecx)
        U^\dagger(\gamma(t))
        O
        U(\gamma(t))
        V(\vecx)
        }{0}^T
    \cdot \nabla_\vectheta \expval{
        V^\dagger(\vecx')
        U^\dagger(\gamma(t))
        O
        U(\gamma(t))
        V(\vecx')
        }{0} \cdot \gamma'(t) \, dt \label{eq:pathker2}  \\
    & \approx 
    \frac{1}{T}
    \sum_{t=0}^{T-1}
    \nabla_\vectheta \expval{
        V^\dagger(\vecx)
        U^\dagger(\gamma_t)
        O
        U(\gamma_t)
        V(\vecx)
        }{0}^T
    \cdot \nabla_\vectheta \expval{
        V^\dagger(\vecx')
        U^\dagger(\gamma_t)
        O
        U(\gamma_t)
        V(\vecx')
        }{0} \label{eq:pathker3}
\end{align}
where Equation \ref{eq:pathker1} defines the QNTK as its classical analog and is equivalent to Equation \ref{eq:pathker2} except for the integration with respect to time. Equation \ref{eq:pathker3} is the discretized version of the preceeding equations, corresponding to  actual implementation in a  gradient descent-trained model. 

The resulting \emph{Quantum Path Kernel} (QPK) is consequently  both a quantized version of Domingo's Path Kernel as well as a generalization of the Quantum NTK, one that is implicitly capable of embodying the complex parametric interactions (such as transient parametric co-evolutions) that occur during learning in order to arrive at the final trained model, including those implicated in hierarchal feature learning.

\subsection{The Quantum Path Kernel as a generalization of Quantum Neural Tangent Kernel}

In interpreting the Quantum Path Kernel as a generalization of QNTK for models exhibiting nonlinear behavior, it may be seen that the QNTK is constant only when independent of $\vectheta$, in which case:
\begin{equation}
    k_\mathrm{qpk}(\vecx, \vecx'; \gamma) = 
    \frac{1}{\lVert \gamma \rVert} 
    \int_\gamma k_\mathrm{qntk}(\vecx, \vecx'; \vectheta) \, d\vectheta
    = k_\mathrm{qntk}(\vecx, \vecx'; \bm{0})
    \int_\gamma \frac{d\vectheta}{\lVert \gamma \rVert}
    = k_\mathrm{qntk}(\vecx, \vecx'; \bm{0}).
\end{equation}

That is, the Quantum Path Kernel becomes identical to the Quantum Neural Tangent Kernel.  
However, as set out in section \ref{sec:background:linearity}, the particular structure of  QNNs will, of itself, give rise to  a nonlinear predictor. Thus, in principle, the QNTK would not be expected to be constant in output terms in the finite width regime \cite{liu2022representation}. However, a close-to-constant behavior can be expected for quantum machine learning models whose training is lazy (i.e. lazy training induced via  overparameterization of the QNN, such that  the large number of parameters result in a simplified loss landscape \cite{peters2022generalization, larocca2021theory}, leading to rapid convergence to a global minima). 

\subsection{Decorrelation in feature representation}\label{sec:quantumpathkernel:decorrelation}

The Quantum Path Kernel clearly exhibits dependency on the training initialization: different initial parameter values, optimization algorithms or learning rates may lead to differing QPK matrices. In particular, the utilization of  `vanilla' gradient-descent optimization algorithms, with a fixed number of training epochs, may introduced subtle biases in the QPK. For example, if training were to converge rapidly, any contribution between the instance of convergence and the end of the training  will be effectively identical and oversampled: this contribution will hence outweight the others, biasing  the `stack' of aggregated kernel matrices toward its final layer, as per~\ref{fig:scheme_path}.

To avoid this, more sophisticated optimization algorithms can be considered. For example, the ADAM optimizer adaptively increases the learning rate in locally convex portions of the loss landscape, leading to fewer similar contributions within the path kernel. Furthermore, it is possible to perturb parameter paths via stochastic, noisy or non-gradient-descent-based optimization techniques in order to decorrelate subsequent contributions to the QPK. Having different, highly decorrelated contributions would allow us to interpret the QPK as an ensemble technique analogous to  bootstrap aggregation (bagging) often used for tuning  the bias/variance trade off in classical machine learning. (Multiple Kernel Learning  \cite{gonen2011multiple} might also be used to optimally weight individual contributions over the kernel at the expense of interpretability in path terms) . 

Appendix \ref{apx:pk:implementation} discuss implementation details for the QPK and its  tested variants. We therefore now turn to an examination of the test regime. 

\section{Experimental evaluation of the Quantum Path Kernel in classifying Gaussian XOR Mixtures}\label{sec:experiments}

\begin{figure*}[htbp]
    \centering
    \includegraphics[width=0.9\textwidth]{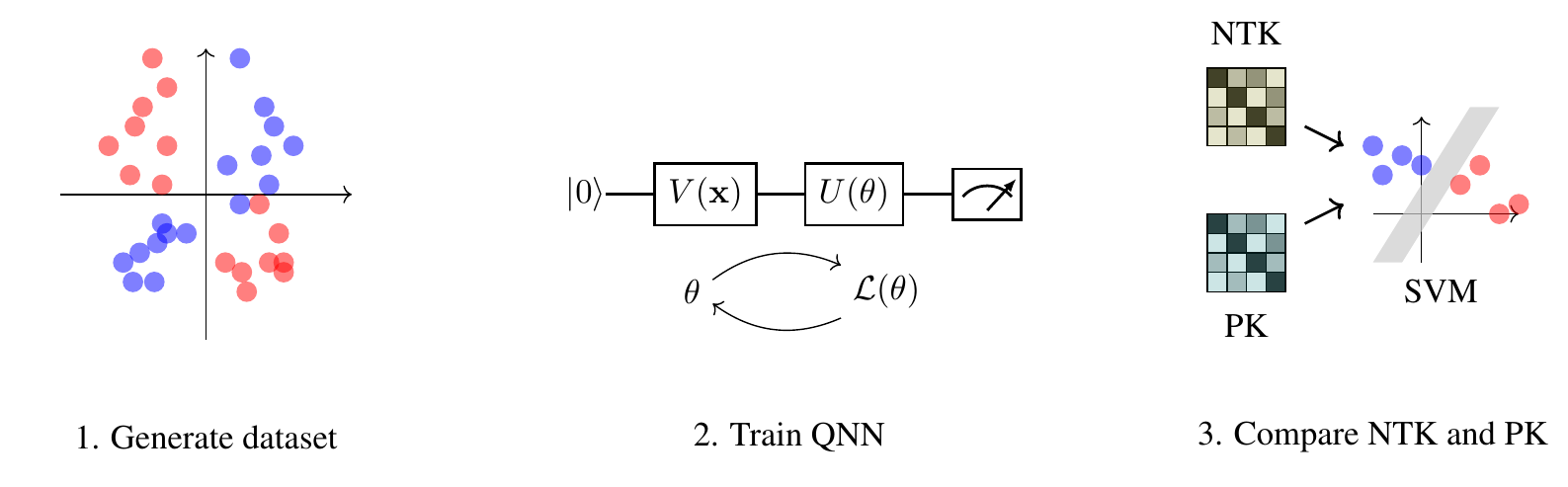}
    \caption{Gaussian XOR Mixture classification experiment workflow.}
    \label{fig:workflow}
\end{figure*}
 
Machine-learning non-linearities such as those underpinning feature learning in empirical DNNs can thus be feasibly implemented in a quantum setting via the QPK. It remains to demonstrate that this can yield superior generalization performance on plausible quantum devices.  Our evaluation, therefore, considers the reference case of the Gaussian XOR Mixture classification problem \cite{deng2019model, mai2019high, lelarge2019asymptotic}.

In particular, the Gaussian XOR Mixture classification problem is an important benchmark for highlighting layer-wise learning capabilities of a model (or the lack of them), in that it intrinsically requires a two-layer solution in order to achieve Bayes optimal class separation. Theoretical evidence has shown that kernel methods, in particular those with random features, struggle to accurately classify XOR data vector mixtures \cite{refinetti2021classifying}. In Appendix~\ref{apx:refinetti} we further analyze the problem, reproducing the results of \cite{refinetti2021classifying}, and proposing an interpretation of the success of feature learning models in tackling the Gaussian XOR Mixture problem.

Our experimental workflow is pictured in Figure~\ref{fig:workflow}. Firstly, we generate the dataset for the above described problem. Secondly, we train several QNNs to best fit the generated data. Thirdly, we use the training information to create the QNTK and QPK matrices; the latter are used to train a kernel machine (specifically the Support Vector Machine) to obtain final classifications. Then, our analysis begins with convergence study of the QNNs with an increasing number of layers, to highlight the effect of architectural parametrization in QNNs. Finally, we compare the performances of the QNTK and QPK approaches in terms of testing and training accuracy. The simulation details are shown in  Appendix~\ref{appendix:simulation}.

\subsection{Experimental Setup}\label{sec:experiment:setup}

\begin{figure*}[tbp]
    \centering
    \includegraphics[width=0.8\textwidth]{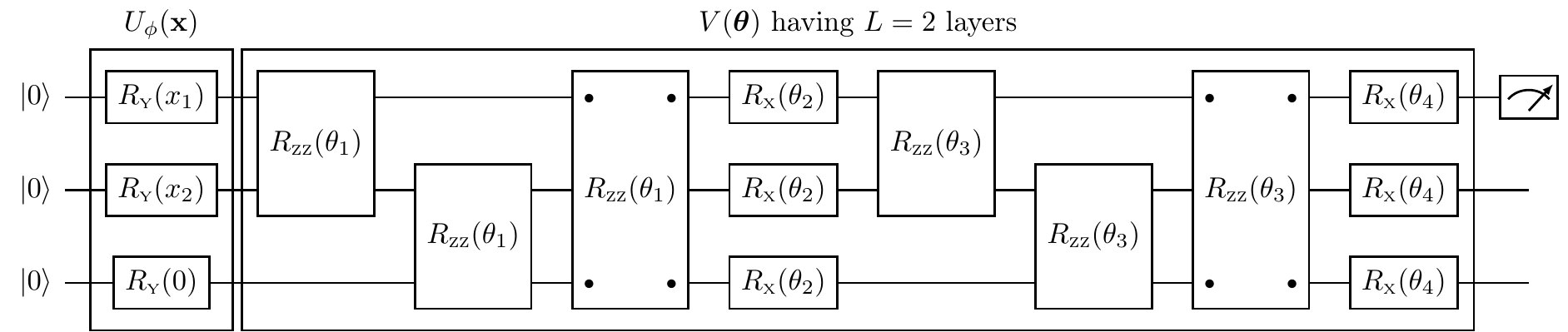}
    \caption{Quantum circuit schematic of the classification model used for $d=3$ qubits and $L=2$ layers. }
    \label{fig:q-circuit}
\end{figure*}

The ground truth Gaussian XOR Mixture dataset is specified by $d$ the dimensionality of the features, $d' \le d$ the number of non-zero features representing the multidimensional Gaussian XOR Mixture, $\bar\epsilon$  the variance of the Gaussian noise, and $n$ the number of data points; it is composed as follows:
\begin{equation}
    \D_{d, d', \bar\epsilon, n} = \left\{
        \begin{array}{r}
           ([x_1+\epsilon_1, ..., x_{d'}+\epsilon_{d'}, 0, ..., 0]^T, y_i) \\
           \in \R^d \times \{\pm 1\}
        \end{array}
    \right\}_{j=1}^n
\end{equation}
where $x_i \sim \{\pm 1\}$, $\epsilon_i \sim \mathcal{N}(0, \bar\epsilon)$ for $i = 1, ..., d'$, and $y_i = \prod_{i=1}^{d'} x_i$. Such a dataset is optimally classified via the oracle function 
\begin{equation}
    f_\mathrm{oracle}(\vecx) = \prod_{i=1}^{d'} x_i.
\end{equation}

We generate multiple datasets $\D_{d, d', \epsilon, n}$ having feature dimensionality ranging in $d = 2, 3, ..., 10$, noise ranging in $\epsilon = 0.1, 0.2, ..., 1.0$, number of non-zero features fixed to $d' = 2$, and number of elements fixed at $n = 32$. Then, each dataset has been randomly partitioned into a training set $\D_\text{train}$ and a testing set $\D_\text{test}$. 

Each dataset is processed by a distinct quantum neural network, each sharing the same structure described by:
\begin{equation}
    f(\vecx; \vectheta)
    = \Trace[\rho_{\vecx, \vectheta} O] 
    = \Trace[V^\dagger(\vectheta) U_\phi^\dagger(\vecx)\rho_0 U_\phi(\vecx) V(\vectheta) O]
\end{equation}
with data encoding:
\begin{equation}
    U_\phi(\vecx) = 
    \prod_{j=1}^{d} 
    \exp{-i \, x_j \sigma_y^{(i)}}
\end{equation}
such that the  trainable ansatz is described:
\begin{equation}
    V(\vectheta)
    = \prod_{j=1}^{L} 
    \exp{-i \, \theta_{2i+1} \sigma_x^{(j)}}
    \exp{-i \, \theta_{2i} \sigma_z^{(j)} \otimes \sigma_z^{(j+1 \text{ mod } d)}}
\end{equation}
with the  $L$ hyperparameter representing the number of layers of the model. Finally, the observable is $O = \sigma_z^{(0)}$. 

This data encoding is been chosen for its simplicity: the encoding of one feature for each qubit results in a constant-depth circuit. The choice of the trainable ansatz, though, is particularly important: the underlying functional transformation has the potential to be affected by barren plateau issues if it is too expressive \cite{holmes2021connecting}, for example when the parametric transformation is able to approximate any arbitrary unitary matrix. The expressibility of a quantum transformation can be examined using Lie-algebraic tools as shown in \cite{larocca2021diagnosing}. Among the class of unitaries that are non-maximally expressive, we have selected a specific form that has empirically demonstrated favorable trainability  as detailed in \cite[Fig. 7a]{larocca2021theory}. The choice of the observable is also guided by the necessity of avoiding the barren plateau issue. According to \cite{cerezo2021cost}, global observables are likely to exhibit vanishing gradients; we thus  apply the simplest possible classifier observable acting on a single qubit. The circuit is pictured in Figure~\ref{fig:q-circuit}. In our experiment, the observed qubit is the uppermost; although any other qubit choice would result in a similar predictor due to the symmetric structure of the circuit.

Each dataset is processed with the above described QNN employing a number of layers ranging from $L = 1$ to $20$. According to \cite{grant2019initialization}, the QNNs should be initialized at $\vectheta = \bm{0}$ to avoid further trainability issues. However, we do not need to consider such initialization strategy for the variational unitary since the previous expedients were sufficient to allow successful training. Thus, the parameters $\theta_j$ are sampled from a standard normal distribution. Each QNN is trained using the stochastic gradient-descent algorithm ADAM for $1000$ epochs using an initial (adaptive) learning rate $ \eta = 0.1$. The loss function is either BCE or MSE and, for the sake of simplification, the batch size is equal to the total cardinality of the training set.

In the experimental setup described above, we study, both epoch-wise and depth-wise, the effect induced by different initialization parameters on the convergence of the loss function during training.

\subsection{Results}

We evaluate the depthwise convergence characteristics of the respective  $f(\vecx; \vectheta)$ models in terms of the corresponding accuracies of the Quantum Path Kernel and Quantum NTK under SVM final classification. Of particular interest is evaluating the closeness of  models to the \emph{lazy training} regime, indicative of the model being near to linear. Lazy training, in classical machine learning, typically occurs for very wide neural networks with the loss decreasing to zero exponentially rapidly, while network parameters stay close to their initialization values throughout  training. In the current context, this would correspond to the  Quantum Path Kernel collapsing to the Quantum Neural Tangent Kernel, and we would anticipate convergent classification performances for the two approaches. 

We therefore evaluate training loss for each of the QNN models over the respective training epochs with an increasing number of QNN layers $L = 1, ..., 20$. This will be used to determine proximity to the lazy training regime (i.e. identifying if the QNN converges exponentially fast to zero loss). We additionally plot the norm difference between the parameters during training compared to their initialization values. These will be used to determine the extend to which parameters vary from their initialization, indicative the training richness of models in the  \emph{non}-lazy training regime. 

We are also interested in determining the robustness of the classifiers to  stochastic noise influences during training and their corresponding resilience to overfitting (or the extent to which \emph{benign overparameterization} \cite{peters2022generalization} effects exists), measured in terms of generalization performance. Therefore, the above evaluations are repeated for datasets additively noise-perturbed in an increasing signal-to-noise ratio. 

Finally, we are interested in comparing the generalization performances of our approach to that of the QNTK. For this, we evaluate  test accuracy score for the QPK and QNTK, against the oracle. Superior performance of the QPK, in solving the Gaussian XOR Mixture problem, will be taken to be indicative of superior ability to replicate the layerwise feature-learning capability of classical multilayer networks.

\subsubsection{Depthwise convergence characteristics}

\newcommand{\figwid}[0]{0.3}

\begin{figure*}[tbp]
\centering
% ================================
\begin{subfigure}[b]{\figwid\textwidth}
\centering
\includegraphics[width=\textwidth]{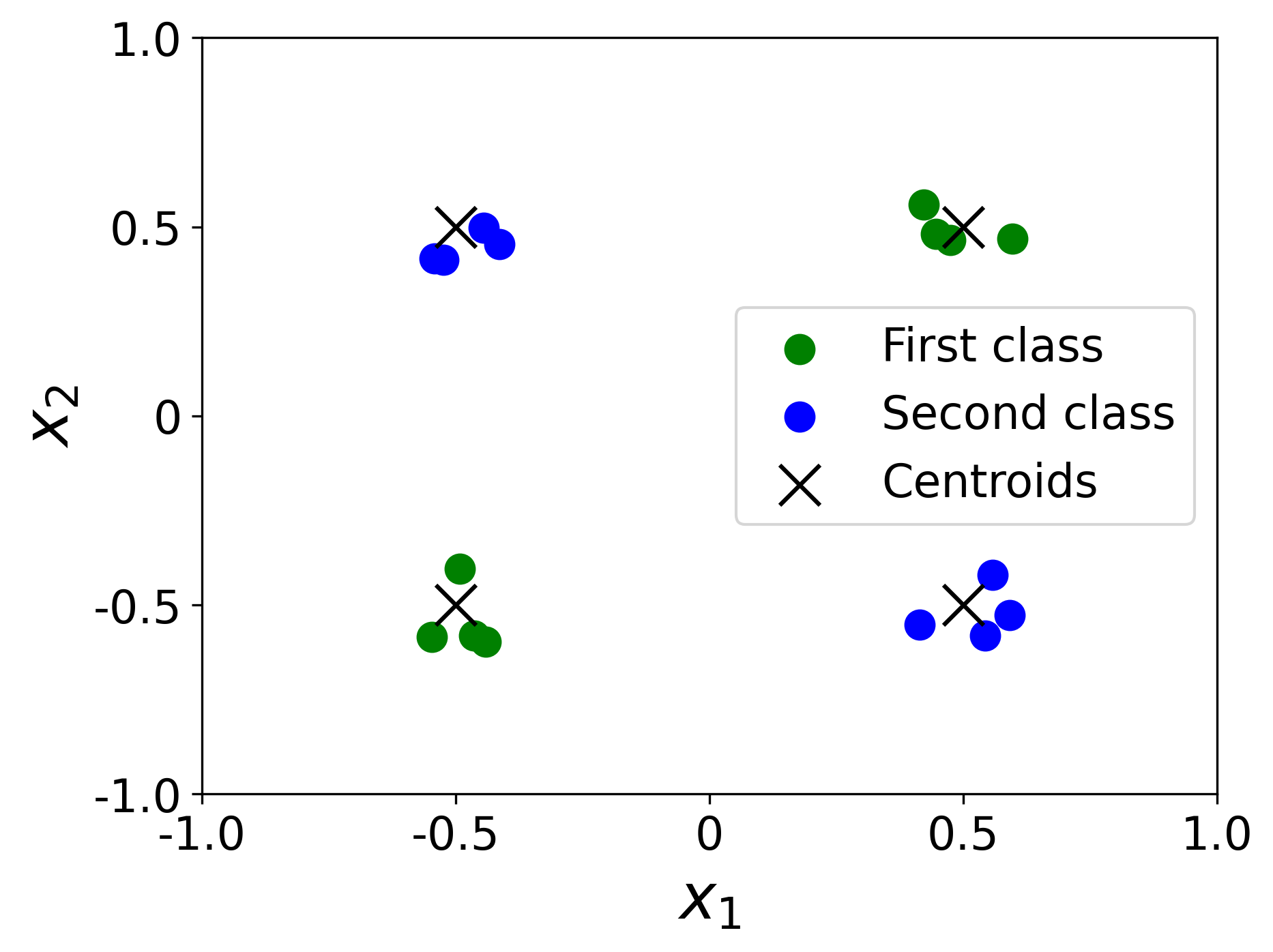}
\caption{}
\label{fig:n01_dataset}
\end{subfigure}
% ================================
\hfill
\begin{subfigure}[b]{\figwid\textwidth}
\centering
\includegraphics[width=\textwidth]{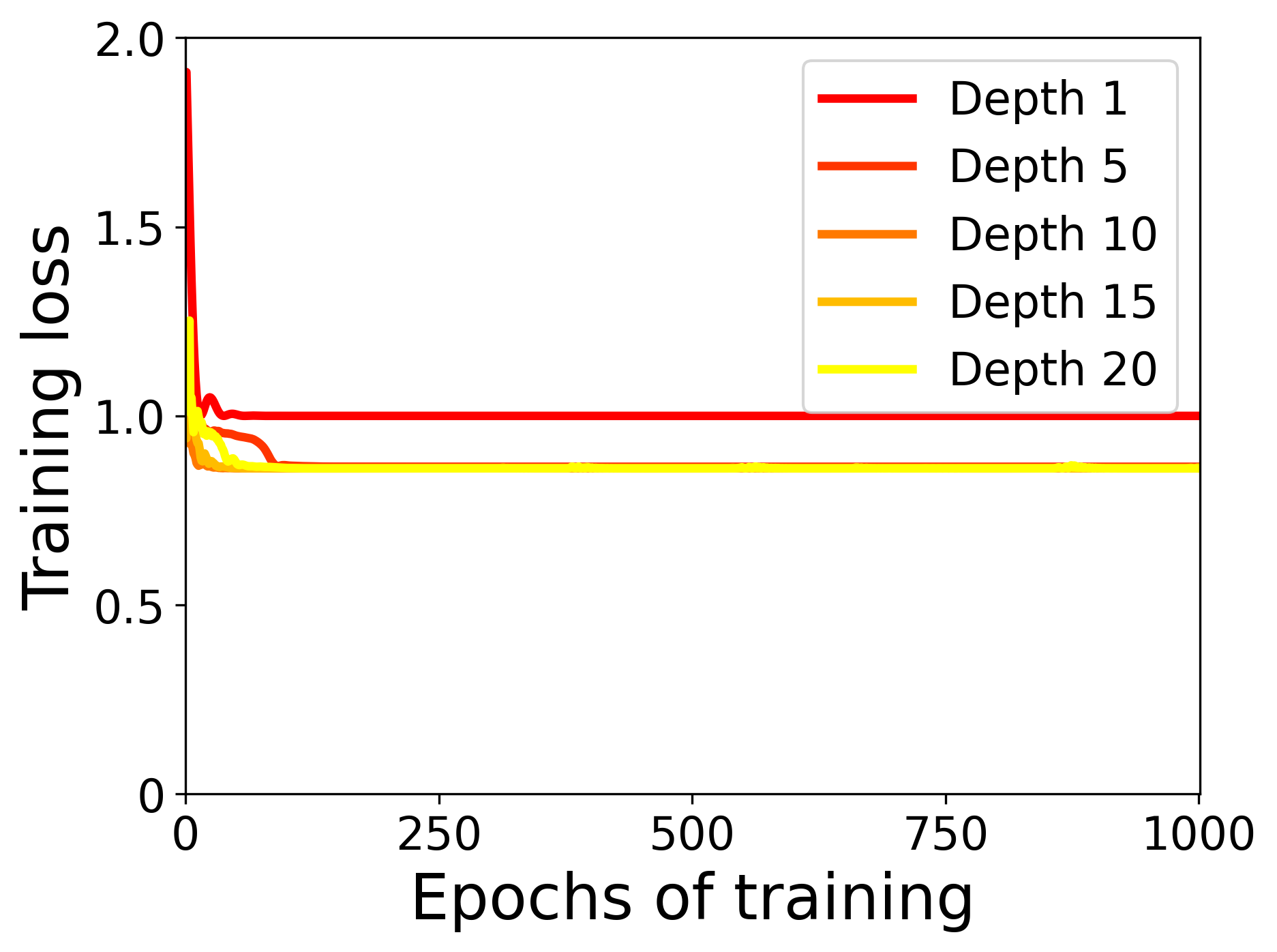}
\caption{}
\label{fig:n01_loss}
\end{subfigure}
% ================================
\hfill
\begin{subfigure}[b]{\figwid\textwidth}
\centering
\includegraphics[width=\textwidth]{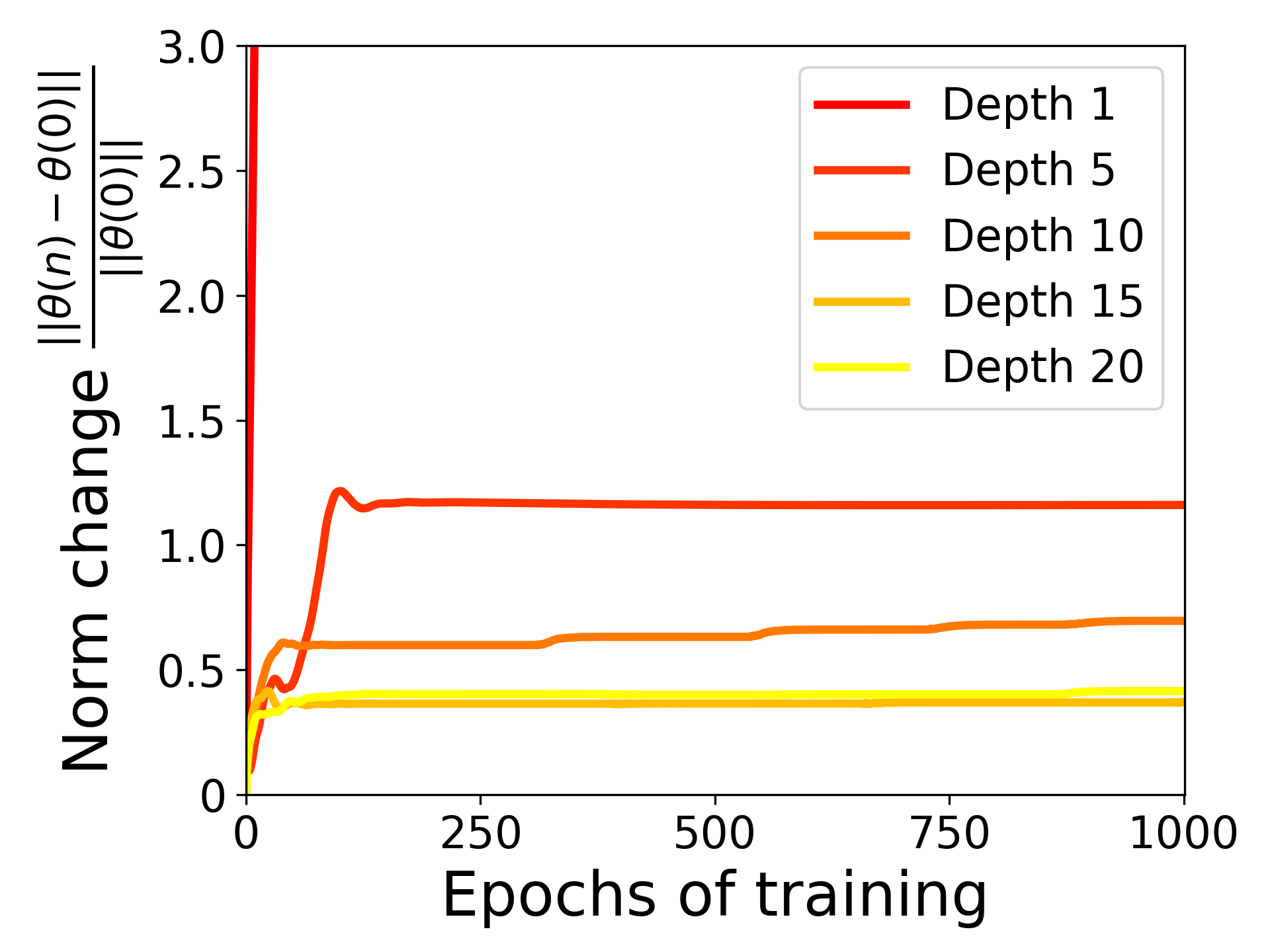}
\caption{}
\label{fig:n01_param}
\end{subfigure}
% ================================
% ================================
\begin{subfigure}[b]{\figwid\textwidth}
\centering
\includegraphics[width=\textwidth]{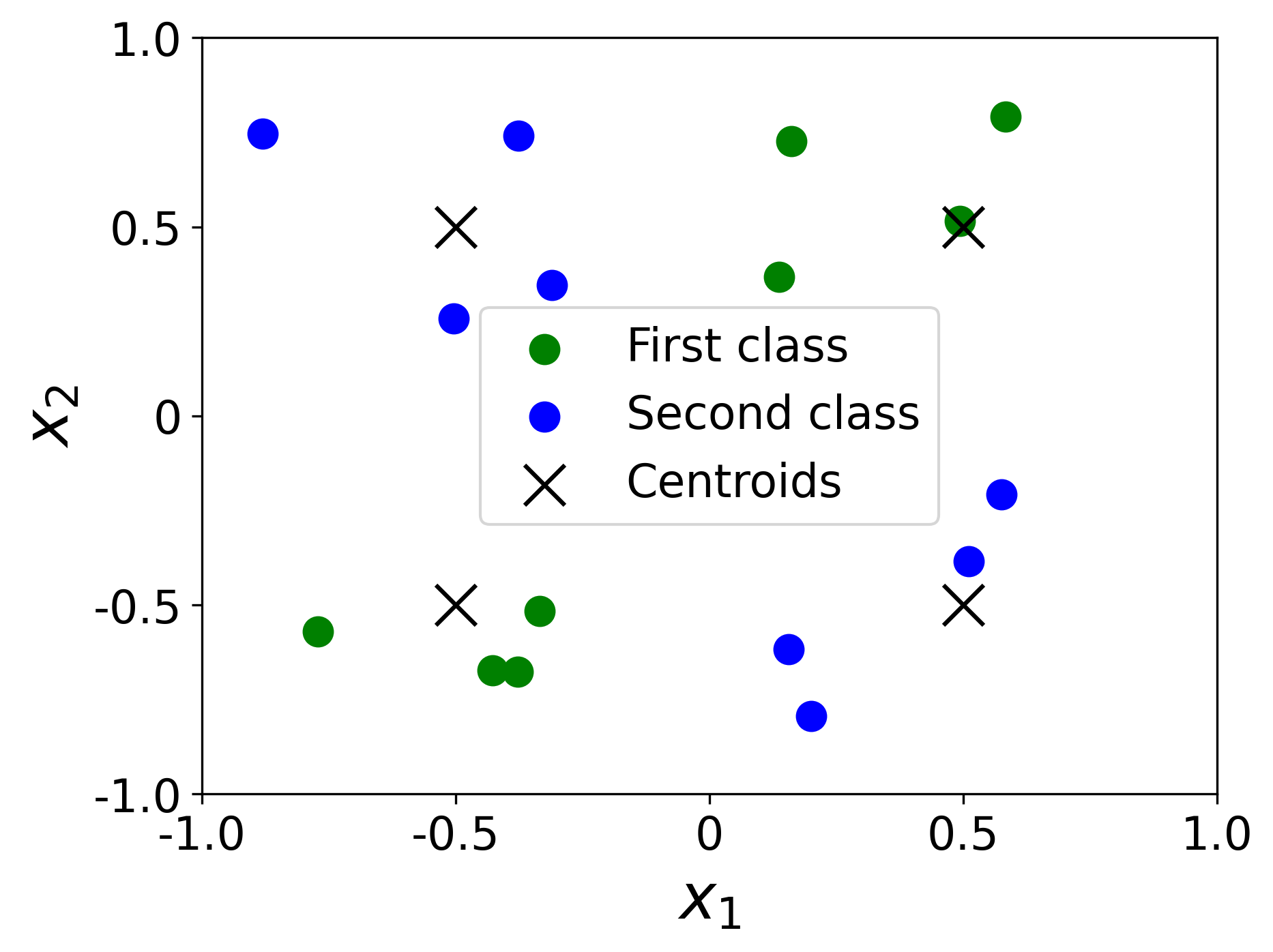}
\caption{}
\label{fig:n04_dataset}
\end{subfigure}
% ================================
\hfill
\begin{subfigure}[b]{\figwid\textwidth}
\centering
\includegraphics[width=\textwidth]{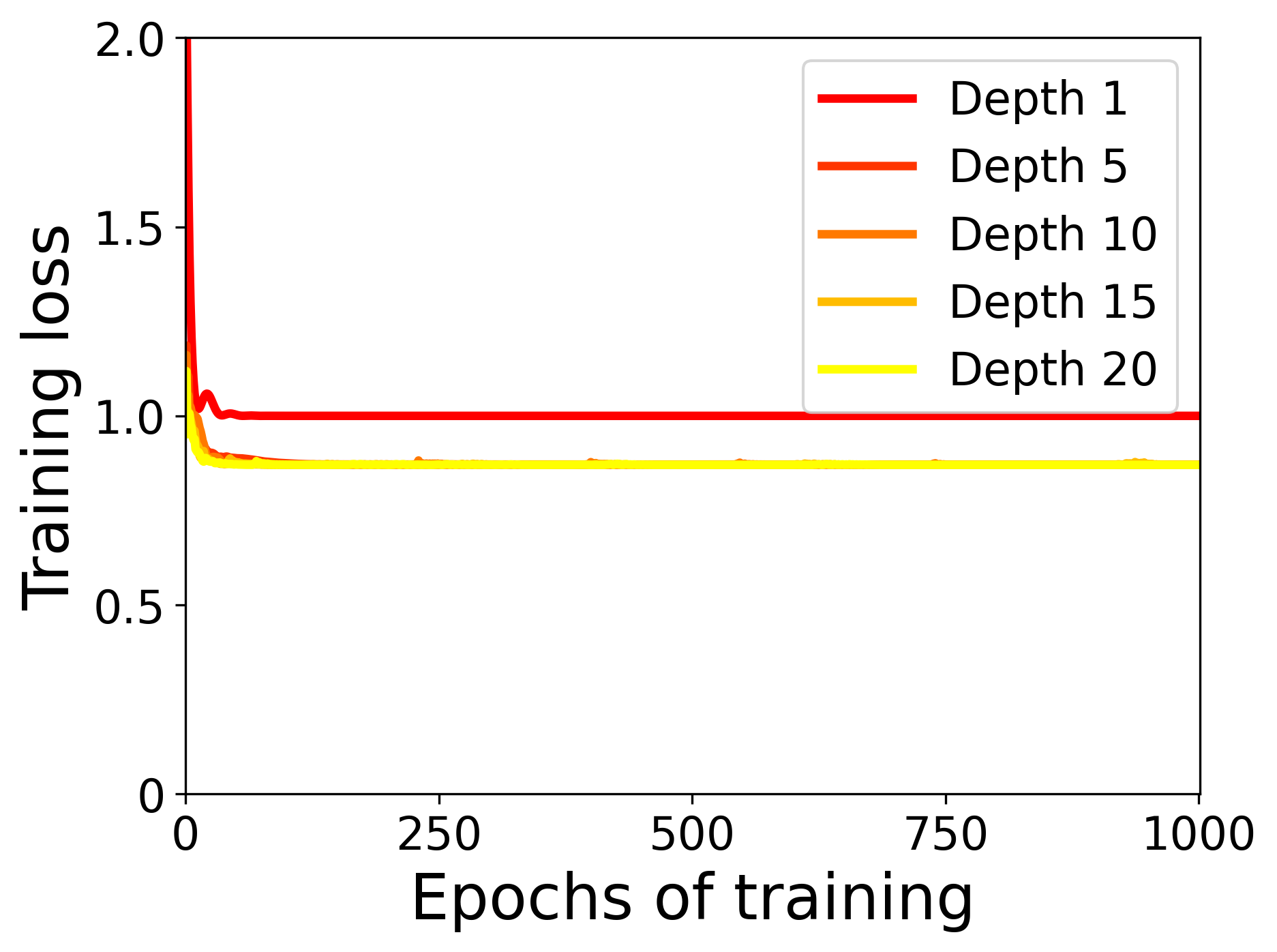}
\caption{}
\label{fig:n04_loss}
\end{subfigure}
% ================================
\hfill
\begin{subfigure}[b]{\figwid\textwidth}
\centering
\includegraphics[width=\textwidth]{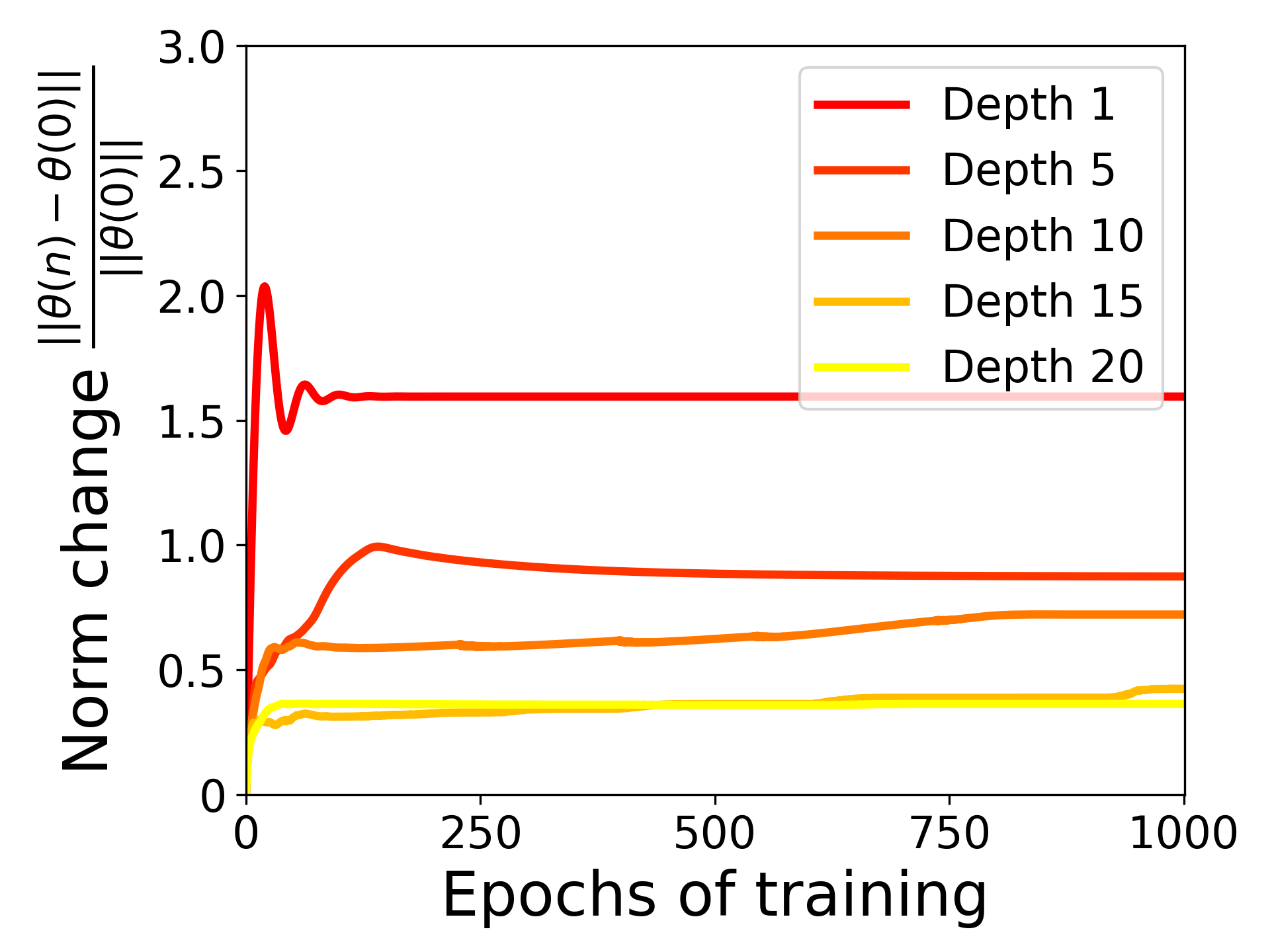}
\caption{}
\label{fig:n04_param}
\end{subfigure}
% ================================
% ================================
\begin{subfigure}[b]{\figwid\textwidth}
\centering
\includegraphics[width=\textwidth]{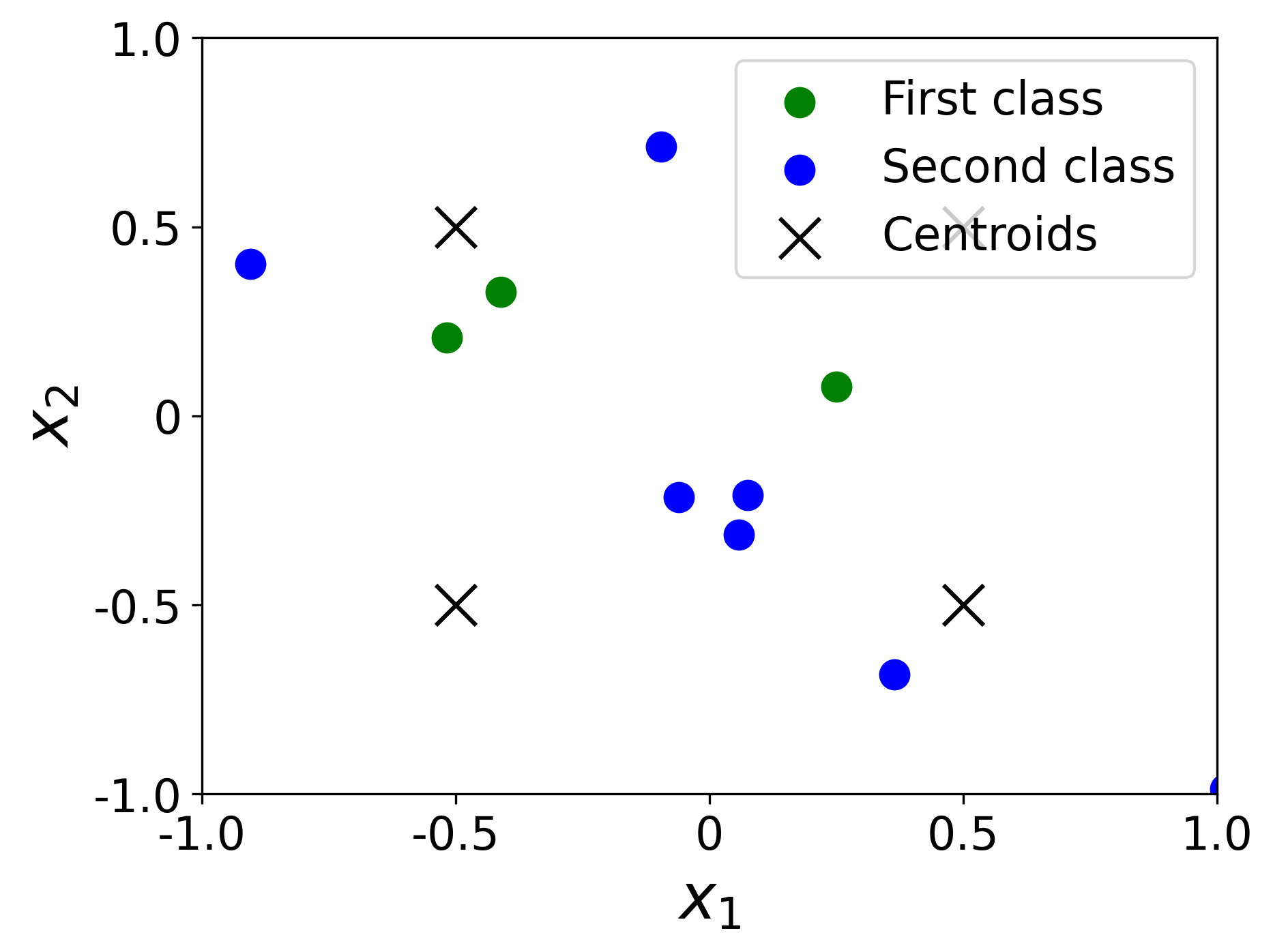}
\caption{}
\label{fig:n10_dataset}
\end{subfigure}
% ================================
\hfill
\begin{subfigure}[b]{\figwid\textwidth}
\centering
\includegraphics[width=\textwidth]{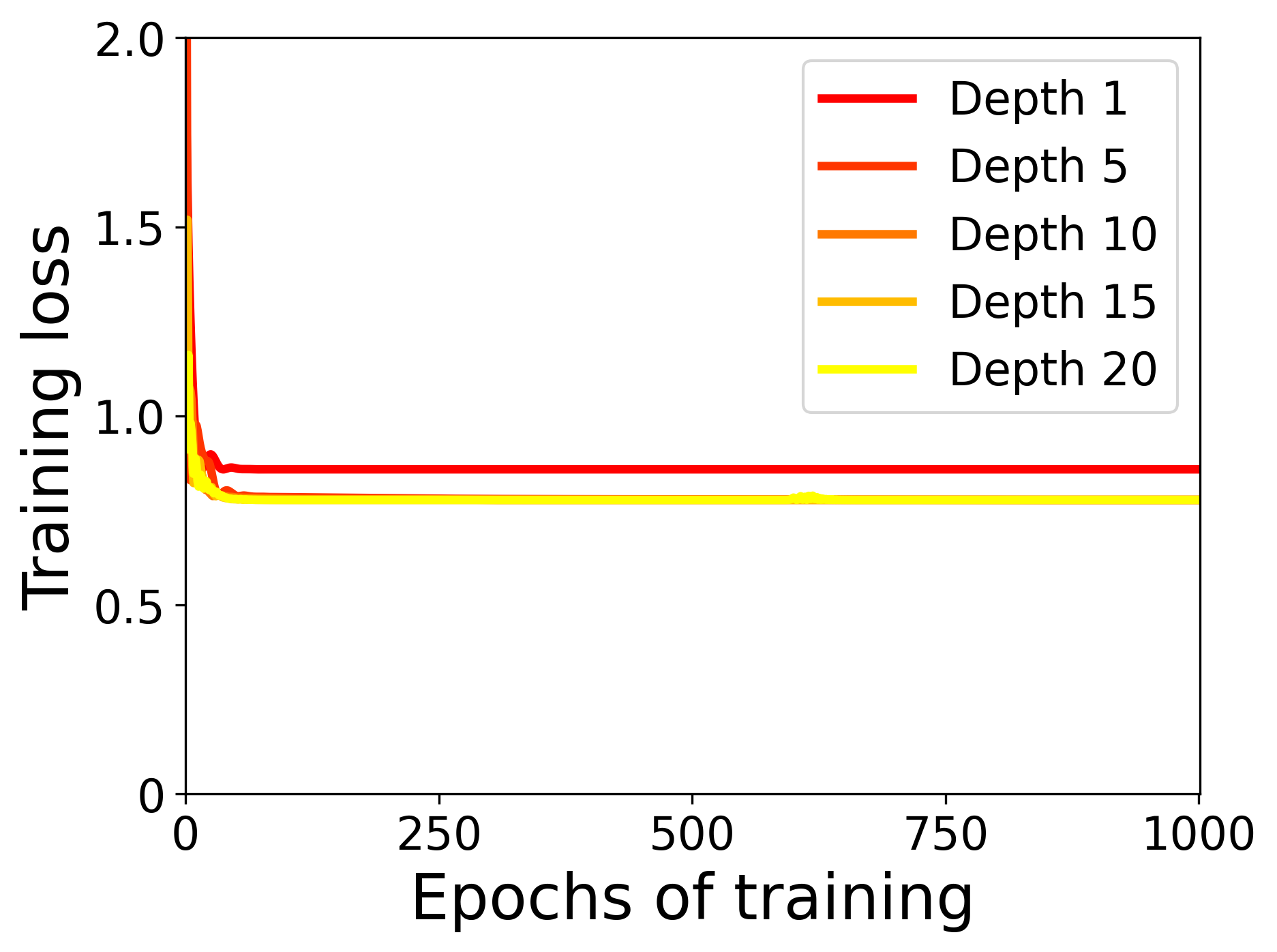}
\caption{}
\label{fig:n10_loss}
\end{subfigure}
% ================================
\hfill
\begin{subfigure}[b]{\figwid\textwidth}
\centering
\includegraphics[width=\textwidth]{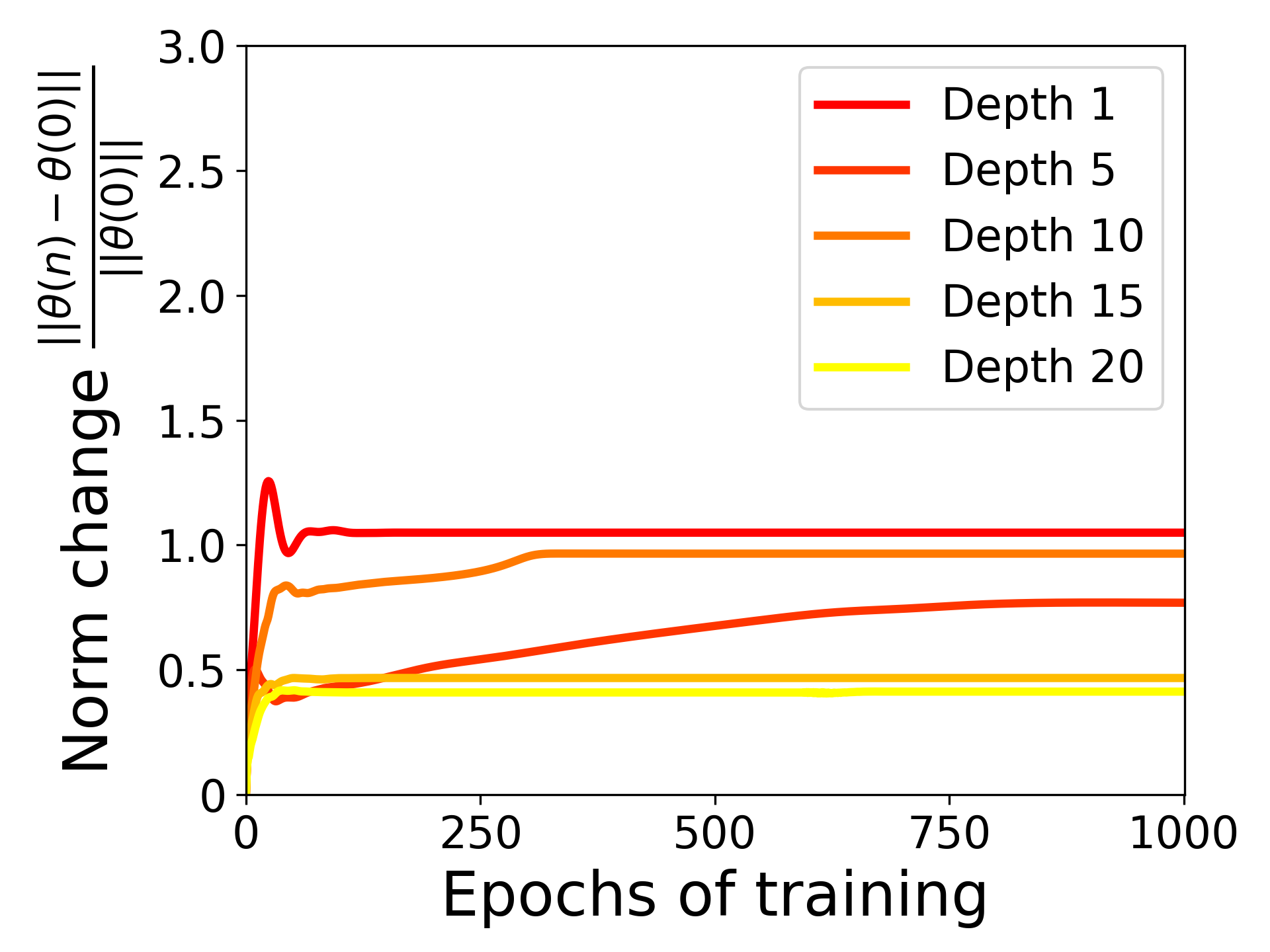}
\caption{}
\label{fig:n10_param}
\end{subfigure}
% ================================
\caption{Behavior of the quantum machine learning models $f(\vecx; \vectheta)$ over the training phase. (\ref{fig:n01_dataset}) illustrates  the training dataset for the parameter selection $d=4, \epsilon=0.1$; (\ref{fig:n01_loss}) shows the evolving loss for each of the 20 evaluated depthwise models ($L=1, ..., 20$) during  training; (\ref{fig:n01_param}) quantifies the deviation of the parameter vector from its initialization. (\ref{fig:n04_dataset}-\ref{fig:n04_loss}-\ref{fig:n04_param}) show the corresponding  information when $d=4, \epsilon=0.4$;  (\ref{fig:n10_dataset}-\ref{fig:n10_loss}-\ref{fig:n10_param}) for $d=4, \epsilon=1.0$.}
\label{fig:training_plots}
\end{figure*}

Figure~\ref{fig:training_plots} indicates the respective convergence behavior of the evaluated quantum machine learning models with respect to the increasing number of layers. Column 1 has illustrative samples from the training distribution with row-wise decrements in the signal-to-noise ratio, column 2 gives the corresponding loss curves during training, and column 3 indicates the corresponding change in the magnitude of the parameter vector offset from initialization:
\begin{equation}
    \frac{\lVert \vectheta(n) - \vectheta(0) \rVert}{\lVert \vectheta(0) \rVert}
\end{equation}
where $\vectheta(0)$ is the value of the parameters at their initialization, and $\vectheta(n)$ is their value at the $n$-th epoch.

It is evident that none of the models reach the interpolation threshold \cite{belkin2018reconciling} - i.e. the point at which the training data is fitted perfectly with zero training error. To fit the training dataset we would need at least 32 parameters (2 non-zero coordinates per point per 16 points). However, we are not able to reach the interpolation threshold even in the deepest configuration with a total of 40 parameters. This behaviour is expected by the choice of a parametrically-constrained $U$ in effect acting as a form of regularisation. As in the classical DNN case, an increasing number of parameters results in a decrease in the loss  (Figure~\ref{fig:n01_loss}-\ref{fig:n04_loss}-\ref{fig:n10_loss}), and in an increase in the proximity between the parameter vectors and their initialization (Figure~\ref{fig:n01_param}-\ref{fig:n04_param}-\ref{fig:n10_param}). 

We can conclude that none of the QNN models exhibit  evidence of lazy training. In particular, while models having a higher number of parameters do indeed converge more rapidly, parameters are nonetheless varying substantially from their initialization. This behaviour is even more noticeable in the smaller models, with a norm difference oscillating substantially prior to  the convergence. Such non-trivial training is suggestive of the QPK differing largely from the QNTK in its training characteristics. 

\subsubsection{Test and train accuracy of the Quantum Path Kernel verses the Quantum NTK}

\begin{figure*}[tbp]
\centering
% ================================
\begin{subfigure}[b]{\figwid\textwidth}
\centering
\includegraphics[width=\textwidth,,trim=0.4cm 0.2cm 0.4cm 0.1cm,clip]{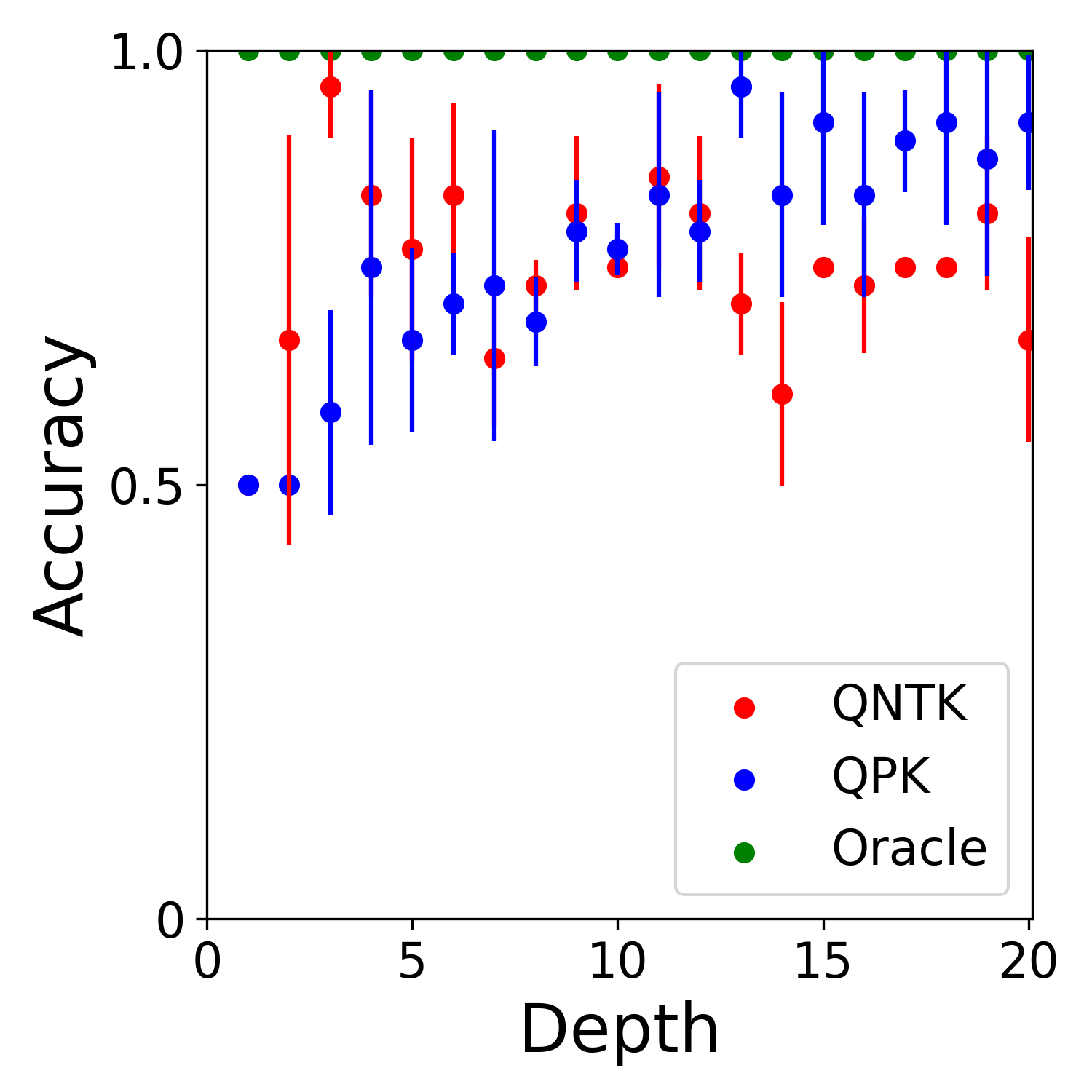}
\caption{}
\label{fig:n01_generr}
\end{subfigure}
% ================================
\hfill
\begin{subfigure}[b]{\figwid\textwidth}
\centering
\includegraphics[width=\textwidth,trim=0.4cm 0.2cm 0.4cm 0.1cm,clip]{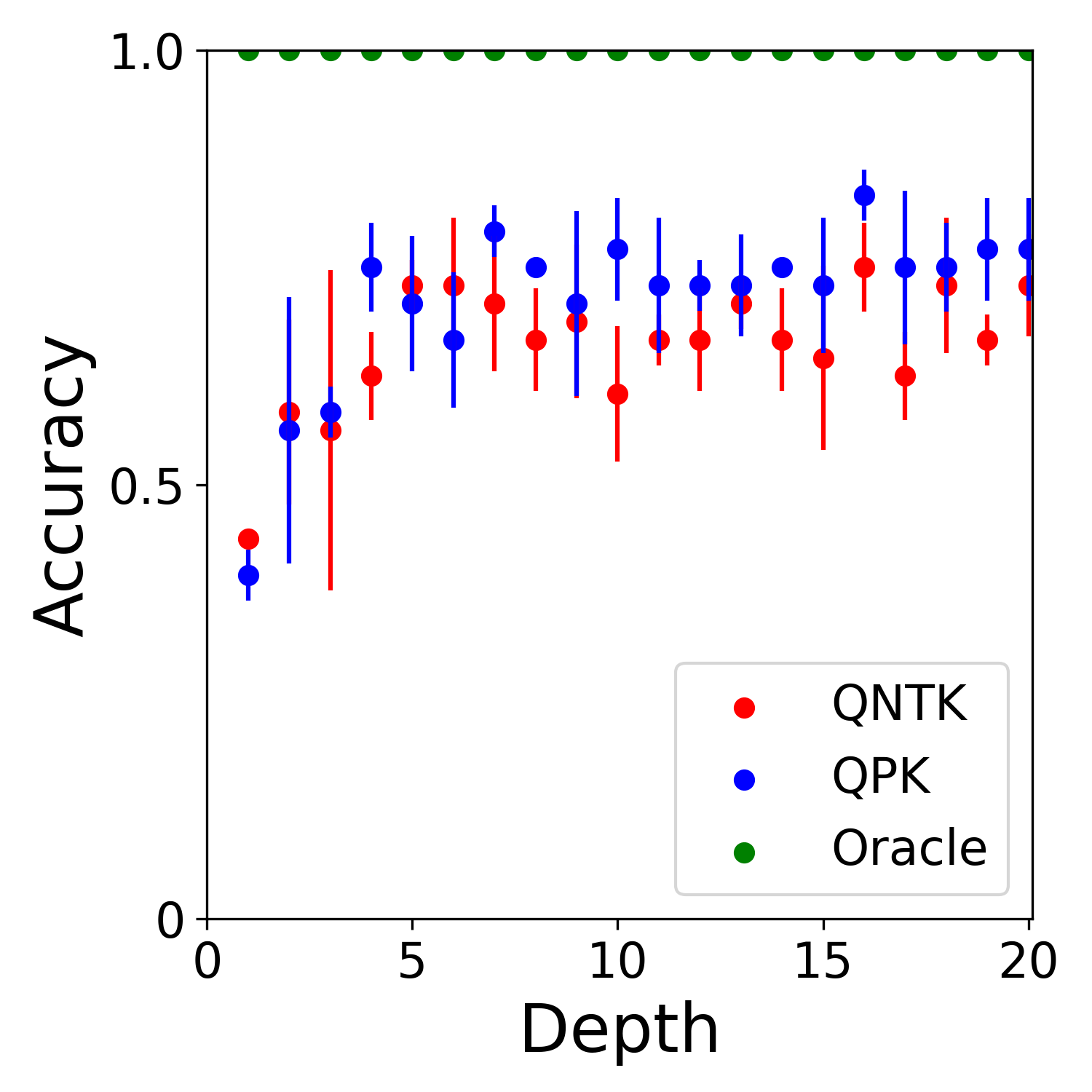}
\caption{}
\label{fig:n04_generr}
\end{subfigure}
% ================================
\hfill
\begin{subfigure}[b]{\figwid\textwidth}
\centering
\includegraphics[width=\textwidth,trim=0.4cm 0.2cm 0.4cm 0.1cm,clip]{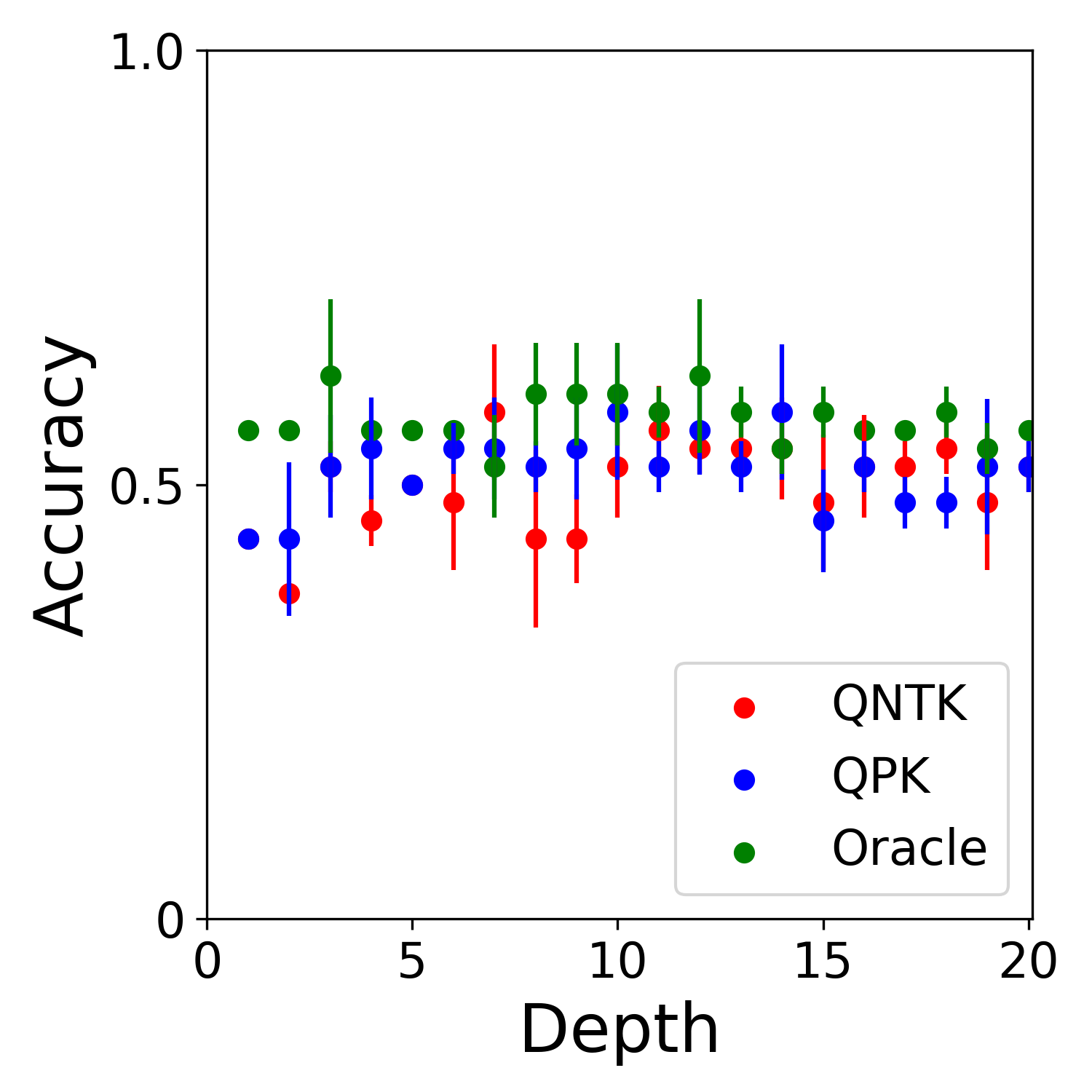}
\caption{}
\label{fig:n10_generr}
\end{subfigure}
% ================================
\caption{Respective test accuracy scores for the Quantum Path Kernel model, the Quantum NTK and the oracle. Error bars represents the standard deviation over three (otherwise identical)  experiments having parametric  specifications  $d=4,\epsilon=0.1$, (\ref{fig:n01_generr}); $d=4, \epsilon=0.4$ (\ref{fig:n04_generr}); $d=4, \epsilon=1.0$ (\ref{fig:n10_generr}).}
\label{fig:accuracy_plots}
\end{figure*}

\begin{figure*}[tbp]
\centering
% ================================
\begin{subfigure}[b]{\figwid\textwidth}
\centering
\includegraphics[width=\textwidth,trim=0.4cm 0.2cm 0.4cm 0.1cm,clip]{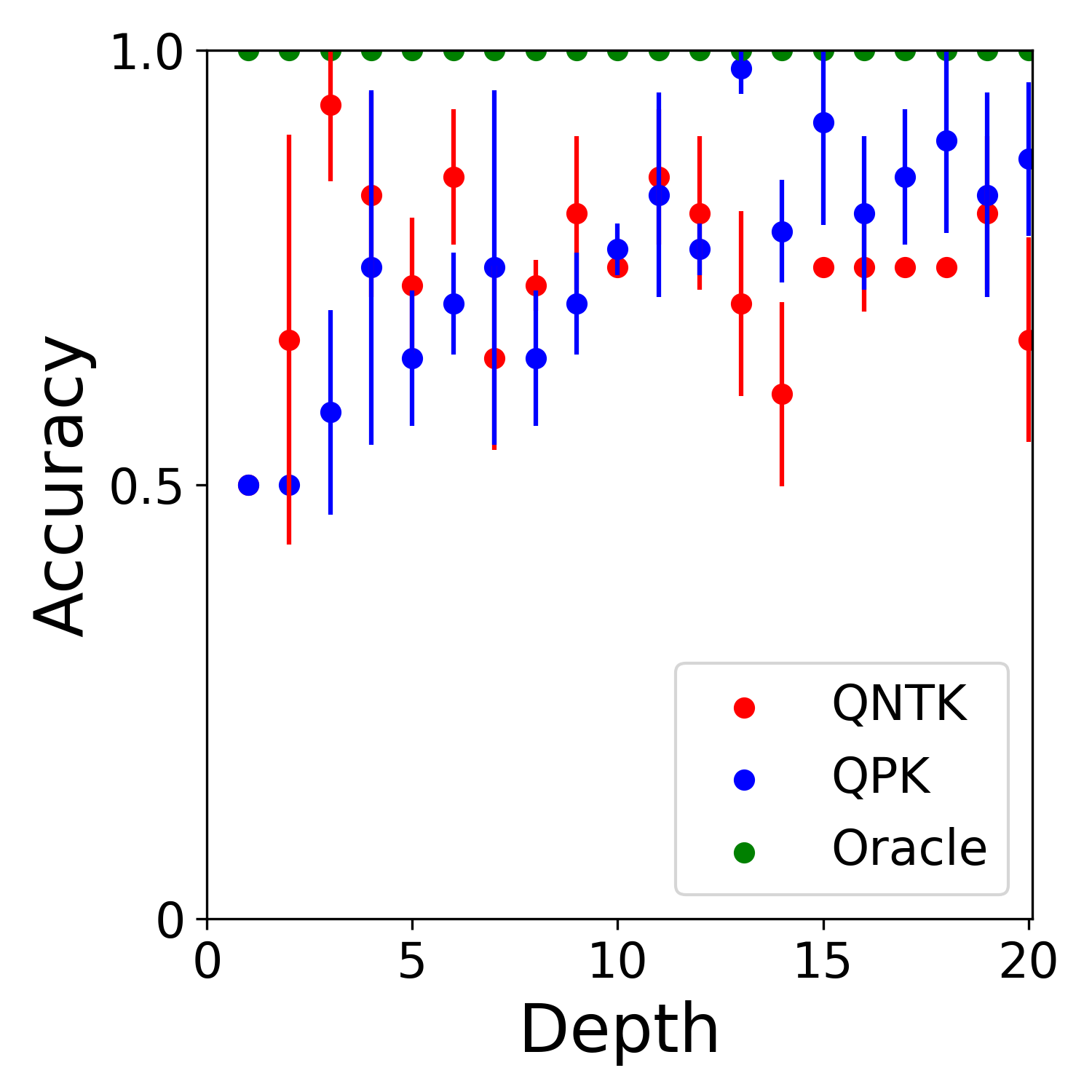}
\caption{}
\label{fig:n01_trainerr}
\end{subfigure}
% ================================
\hfill
\begin{subfigure}[b]{\figwid\textwidth}
\centering
\includegraphics[width=\textwidth,trim=0.4cm 0.2cm 0.4cm 0.1cm,clip]{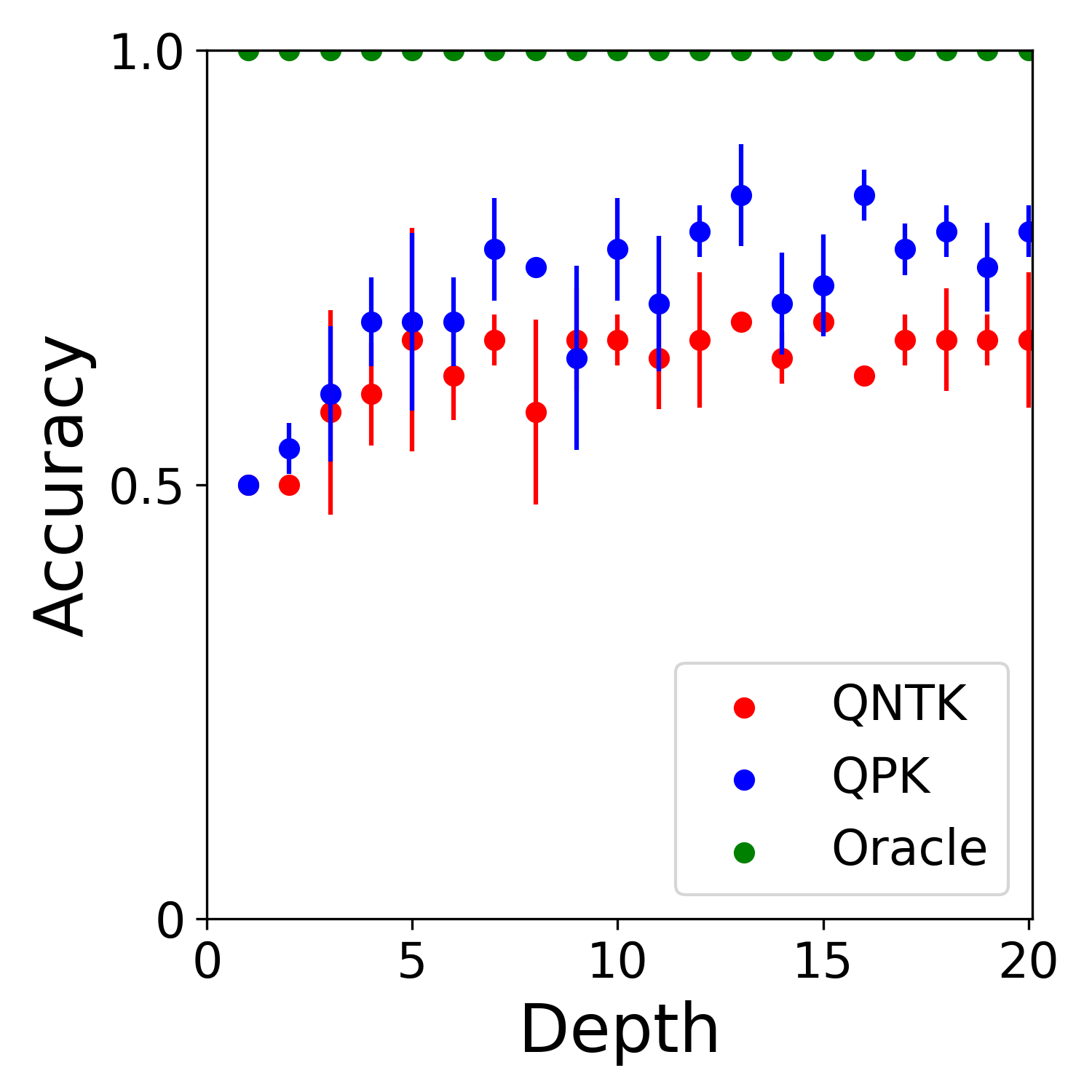}
\caption{}
\label{fig:n04_trainerr}
\end{subfigure}
% ================================
\hfill
\begin{subfigure}[b]{\figwid\textwidth}
\centering
\includegraphics[width=\textwidth,trim=0.4cm 0.2cm 0.4cm 0.1cm,clip]{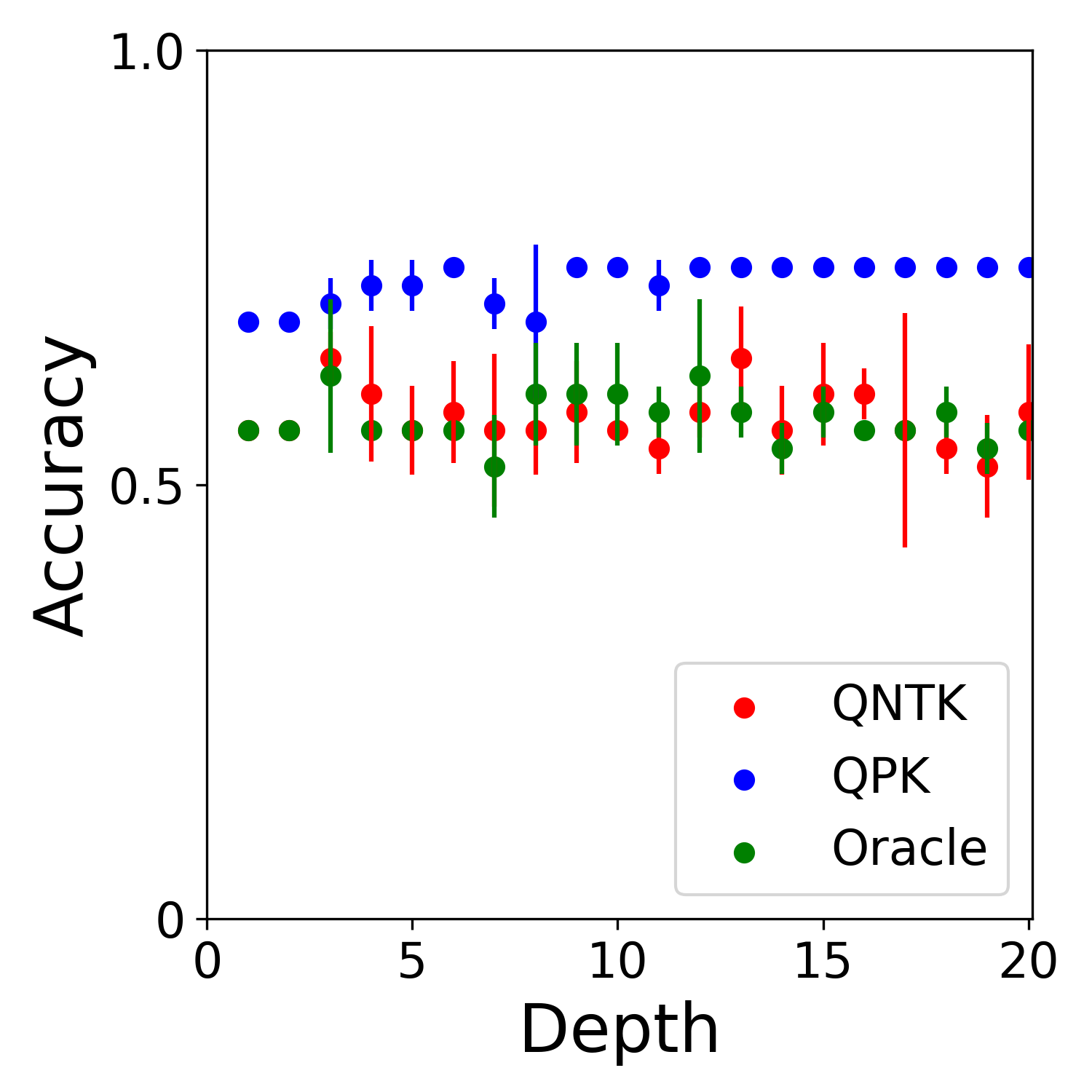}
\caption{}
\label{fig:n10_trainerr}
\end{subfigure}
% ================================
\caption{Respective training accuracies of the Quantum Path Kernel model, the Quantum NTK and the oracle. Error bars represents the standard deviation over three (otherwise identical) experiments having specifications $d=4,\epsilon=0.1$, (\ref{fig:n01_trainerr}); $d=4, \epsilon=0.4$ (\ref{fig:n04_trainerr}); $d=4, \epsilon=1.0$ (\ref{fig:n10_trainerr}).}
\label{fig:accuracy_training_plots}
\end{figure*}

Figure~\ref{fig:accuracy_plots} indicates the corresponding test accuracies, measuring how well the respective models generalize to unseen data. While the QPK and Quantum NTK models both perform similarly at low signal-to-noise ratios, it is particularly striking to observe the outperformance of the QPK over the Quantum NTK  with increasing hierarchical depth at the highest signal-to-noise setting..

Figure~\ref{fig:accuracy_training_plots} indicates the training accuracy with depth at the point of convergence. It may be observed that the QPK exhibits lower loss than the Quantum NTK across the full signal-to-noise range, with the effect becoming more marked at higher noise levels (ultimately over-fitting relative to the noise-free oracle in panel $c$), consistent with the expectation that QPK has a lower bias than the Quantum NTK. 

In sum, results confirm the anticipated improvement in performance for the QPK over the QNTK in the Gaussian XOR mixture setting. 

\section{Conclusion and Further Work}\label{sec:conclusion}

We have introduced the Quantum Path Kernel as a mechanism for incorporating key complex classical multi-layer network learning behaviors, in particular hierarchical feature learning, within quantum neural networks via an appropriately expressive kernelization of the training process. We evaluate our approach on the Gaussian XOR mixture classification problem, a straightforward benchmark of multilayer learning capacity that requires a minimum two-layer solution in order to approach Bayes optimally. Experimental results indicate  superior generalization performance relative to the Quantum NTK, an advantage which is especially pronounced in high-depth, low signal-to-noise settings. 

We have shown theoretically  that the Quantum Path Kernel converges to the Quantum NTK only in the lazy training regime, i.e. when the training loss decreases to zero exponentially fast whilst  model parameters stay close to their initializations across training. Such  behaviour is classically seen in infinite-wide neural networks, whose behaviour is  then close to that of a linear model. Our experiments, by contrast, indicate that QNNs do not operate in the linear regime.

We have discussed, though do not evaluate in the current paper,  the potential for  using stochastic, noisy or non-gradient descent based optimization techniques to artificially perturb parameter paths within the QPK in order to implicate more decorrelated feature representations. We, furthermore, propose in future to extend the QPK approach via weighting of individual kernel representations in a more heuristic way, for example via  Multiple Kernel Learning. We have also referred in passing to the interpretation of the QPK as an ensemble method due to the averaging operation over its kernel matrices. This will be explored more fully in future investigations. 

\bibliographystyle{unsrt}
%\bibliography{biblio}

\begin{thebibliography}{10}

\bibitem{wittek2014quantum}
Peter Wittek.
\newblock {\em Quantum machine learning: what quantum computing means to data
  mining}.
\newblock Academic Press, 2014.

\bibitem{schuld2015introduction}
Maria Schuld, Ilya Sinayskiy, and Francesco Petruccione.
\newblock An introduction to quantum machine learning.
\newblock {\em Contemporary Physics}, 56(2):172--185, 2015.

\bibitem{huang2021power}
Hsin-Yuan Huang, Michael Broughton, Masoud Mohseni, Ryan Babbush, Sergio Boixo,
  Hartmut Neven, and Jarrod~R McClean.
\newblock Power of data in quantum machine learning.
\newblock {\em Nature communications}, 12(1):1--9, 2021.

\bibitem{abbas2021power}
Amira Abbas, David Sutter, Christa Zoufal, Aur{\'e}lien Lucchi, Alessio
  Figalli, and Stefan Woerner.
\newblock The power of quantum neural networks.
\newblock {\em Nature Computational Science}, 1(6):403--409, 2021.

\bibitem{liu2021rigorous}
Yunchao Liu, Srinivasan Arunachalam, and Kristan Temme.
\newblock A rigorous and robust quantum speed-up in supervised machine
  learning.
\newblock {\em Nature Physics}, 17(9):1013--1017, 2021.

\bibitem{huang2022quantum}
Hsin-Yuan Huang, Michael Broughton, Jordan Cotler, Sitan Chen, Jerry Li, Masoud
  Mohseni, Hartmut Neven, Ryan Babbush, Richard Kueng, John Preskill, et~al.
\newblock Quantum advantage in learning from experiments.
\newblock {\em Science}, 376(6598):1182--1186, 2022.

\bibitem{mcclean2018barren}
Jarrod~R. McClean, Sergio Boixo, Vadim~N. Smelyanskiy, Ryan Babbush, and
  Hartmut Neven.
\newblock Barren plateaus in quantum neural network training landscapes.
\newblock {\em Nature Communications}, 9(1), 11 2018.

\bibitem{holmes2021connecting}
Zo{\"e} Holmes, Kunal Sharma, Marco Cerezo, and Patrick~J Coles.
\newblock Connecting ansatz expressibility to gradient magnitudes and barren
  plateaus.
\newblock {\em PRX Quantum}, 3(1):010313, 2022.

\bibitem{skolik2021layerwise}
Andrea Skolik, Jarrod~R. McClean, Masoud Mohseni, Patrick van~der Smagt, and
  Martin Leib.
\newblock Layerwise learning for quantum neural networks.
\newblock {\em Quantum Machine Intelligence}, 3(1), 1 2021.

\bibitem{schuld2021supervised}
Maria Schuld.
\newblock Supervised quantum machine learning models are kernel methods, 2021.

\bibitem{liu2022representation}
Junyu Liu, Francesco Tacchino, Jennifer~R Glick, Liang Jiang, and Antonio
  Mezzacapo.
\newblock Representation learning via quantum neural tangent kernels.
\newblock {\em PRX Quantum}, 3(3):030323, 2022.

\bibitem{sharma2020trainability}
Kunal Sharma, Marco Cerezo, Lukasz Cincio, and Patrick~J Coles.
\newblock Trainability of dissipative perceptron-based quantum neural networks.
\newblock {\em Physical Review Letters}, 128(18):180505, 2022.

\bibitem{jacot2018neural}
Arthur Jacot, Franck Gabriel, and Cl{\'e}ment Hongler.
\newblock Neural tangent kernel: Convergence and generalization in neural
  networks.
\newblock {\em Advances in neural information processing systems}, 31, 2018.

\bibitem{chizat2019lazy}
Lenaic Chizat, Edouard Oyallon, and Francis Bach.
\newblock On lazy training in differentiable programming.
\newblock {\em Advances in Neural Information Processing Systems}, 32, 2019.

\bibitem{liu2020linearity}
Chaoyue Liu, Libin Zhu, and Misha Belkin.
\newblock On the linearity of large non-linear models: when and why the tangent
  kernel is constant.
\newblock {\em Advances in Neural Information Processing Systems},
  33:15954--15964, 2020.

\bibitem{ghorbani2021neural}
Behrooz Ghorbani, Song Mei, Theodor Misiakiewicz, and Andrea Montanari.
\newblock When do neural networks outperform kernel methods?
\newblock {\em Journal of Statistical Mechanics: Theory and Experiment},
  2021(12):124009, 2021.

\bibitem{chen2020towards}
Minshuo Chen, Yu~Bai, Jason~D Lee, Tuo Zhao, Huan Wang, Caiming Xiong, and
  Richard Socher.
\newblock Towards understanding hierarchical learning: Benefits of neural
  representations.
\newblock {\em Advances in Neural Information Processing Systems},
  33:22134--22145, 2020.

\bibitem{shirai2021quantum}
Norihito Shirai, Kenji Kubo, Kosuke Mitarai, and Keisuke Fujii.
\newblock Quantum tangent kernel.
\newblock {\em arXiv preprint arXiv:2111.02951}, 2021.

\bibitem{domingos2020every}
Pedro Domingos.
\newblock Every model learned by gradient descent is approximately a kernel
  machine.
\newblock {\em arXiv preprint arXiv:2012.00152}, 2020.

\bibitem{refinetti2021classifying}
Maria Refinetti, Sebastian Goldt, Florent Krzakala, and Lenka Zdeborov{\'a}.
\newblock Classifying high-dimensional gaussian mixtures: Where kernel methods
  fail and neural networks succeed.
\newblock In {\em International Conference on Machine Learning}, pages
  8936--8947. PMLR, 2021.

\bibitem{ghorbani2019investigation}
Behrooz Ghorbani, Shankar Krishnan, and Ying Xiao.
\newblock An investigation into neural net optimization via hessian eigenvalue
  density.
\newblock In {\em International Conference on Machine Learning}, pages
  2232--2241. PMLR, 2019.

\bibitem{arora2019exact}
Sanjeev Arora, Simon~S Du, Wei Hu, Zhiyuan Li, Russ~R Salakhutdinov, and
  Ruosong Wang.
\newblock On exact computation with an infinitely wide neural net.
\newblock {\em Advances in Neural Information Processing Systems}, 32, 2019.

\bibitem{bai2020taylorized}
Yu~Bai, Ben Krause, Huan Wang, Caiming Xiong, and Richard Socher.
\newblock Taylorized training: Towards better approximation of neural network
  training at finite width.
\newblock {\em arXiv preprint arXiv:2002.04010}, 2020.

\bibitem{schuld2014quest}
Maria Schuld, Ilya Sinayskiy, and Francesco Petruccione.
\newblock The quest for a quantum neural network.
\newblock {\em Quantum Information Processing}, 13(11):2567--2586, 2014.

\bibitem{schuld2015simulating}
Maria Schuld, Ilya Sinayskiy, and Francesco Petruccione.
\newblock Simulating a perceptron on a quantum computer.
\newblock {\em Physics Letters A}, 379(7):660–663, 3 2015.

\bibitem{cao2017quantum}
Yudong Cao, Gian~Giacomo Guerreschi, and Al{\'a}n Aspuru-Guzik.
\newblock Quantum neuron: an elementary building block for machine learning on
  quantum computers.
\newblock {\em arXiv preprint arXiv:1711.11240}, 2017.

\bibitem{hu2018towards}
Wei Hu.
\newblock Towards a real quantum neuron.
\newblock {\em Natural Science}, 10(3):99--109, 2018.

\bibitem{gili2022introducing}
Kaitlin Gili, Mykolas Sveistrys, and Chris Ballance.
\newblock Introducing non-linearity into quantum generative models.
\newblock {\em arXiv preprint arXiv:2205.14506}, 2022.

\bibitem{tacchino2019artificial}
Francesco Tacchino, Chiara Macchiavello, Dario Gerace, and Daniele Bajoni.
\newblock An artificial neuron implemented on an actual quantum processor.
\newblock {\em npj Quantum Information}, 5(1):1--8, 2019.

\bibitem{sharma2022trainability}
Kunal Sharma, Marco Cerezo, Lukasz Cincio, and Patrick~J Coles.
\newblock Trainability of dissipative perceptron-based quantum neural networks.
\newblock {\em Physical Review Letters}, 128(18):180505, 2022.

\bibitem{guo2021nonlinear}
Naixu Guo, Kosuke Mitarai, and Keisuke Fujii.
\newblock Nonlinear transformation of complex amplitudes via quantum singular
  value transformation.
\newblock {\em arXiv preprint arXiv:2107.10764}, 2021.

\bibitem{holmes2021nonlinear}
Zo{\"e} Holmes, Nolan Coble, Andrew~T Sornborger, and Yi{\u{g}}it
  Suba{\c{s}}{\i}.
\newblock On nonlinear transformations in quantum computation.
\newblock {\em arXiv preprint arXiv:2112.12307}, 2021.

\bibitem{daskin2018simple}
Ammar Daskin.
\newblock A simple quantum neural net with a periodic activation function.
\newblock In {\em 2018 IEEE International Conference on Systems, Man, and
  Cybernetics (SMC)}, pages 2887--2891. IEEE, 2018.

\bibitem{Weinberg1989Precision}
Steven Weinberg.
\newblock Precision tests of quantum mechanics.
\newblock {\em Phys. Rev. Lett.}, 62:485--488, 1 1989.

\bibitem{abrams1998nonlinear}
Daniel~S. Abrams and Seth Lloyd.
\newblock Nonlinear quantum mechanics implies polynomial-time solution for
  $\mathit{NP}$-complete and \# $\mathit{P}$ problems.
\newblock {\em Phys. Rev. Lett.}, 81:3992--3995, 11 1998.

\bibitem{jerbi2021quantum}
Sofiene Jerbi, Lukas~J Fiderer, Hendrik~Poulsen Nautrup, Jonas~M K{\"u}bler,
  Hans~J Briegel, and Vedran Dunjko.
\newblock Quantum machine learning beyond kernel methods.
\newblock {\em arXiv preprint arXiv:2110.13162}, 2021.

\bibitem{roberts2021principles}
Daniel~A. Roberts, Sho Yaida, and Boris Hanin.
\newblock {\em The Principles of Deep Learning Theory}.
\newblock Cambridge University Press, 2022.

\bibitem{lecun2015deep}
Yann LeCun, Yoshua Bengio, and Geoffrey Hinton.
\newblock Deep learning.
\newblock {\em Nature}, 521(7553):436--444, 2015.

\bibitem{peters2022generalization}
Evan Peters and Maria Schuld.
\newblock Generalization despite overfitting in quantum machine learning
  models.
\newblock {\em arXiv preprint arXiv:2209.05523}, 2022.

\bibitem{larocca2021theory}
Martin Larocca, Nathan Ju, Diego García-Martín, Patrick~J. Coles, and
  M.~Cerezo.
\newblock Theory of overparametrization in quantum neural networks, 2021.

\bibitem{gonen2011multiple}
Mehmet G{\"o}nen and Ethem Alpayd{\i}n.
\newblock Multiple kernel learning algorithms.
\newblock {\em The Journal of Machine Learning Research}, 12:2211--2268, 2011.

\bibitem{deng2019model}
Zeyu Deng, Abla Kammoun, and Christos Thrampoulidis.
\newblock A model of double descent for high-dimensional binary linear
  classification.
\newblock {\em arXiv preprint arXiv:1911.05822}, 2019.

\bibitem{mai2019high}
Xiaoyi Mai and Zhenyu Liao.
\newblock High dimensional classification via regularized and unregularized
  empirical risk minimization: Precise error and optimal loss.
\newblock {\em arXiv preprint arXiv:1905.13742}, 2019.

\bibitem{lelarge2019asymptotic}
Marc Lelarge and L{\'e}o Miolane.
\newblock Asymptotic bayes risk for gaussian mixture in a semi-supervised
  setting.
\newblock In {\em 2019 IEEE 8th International Workshop on Computational
  Advances in Multi-Sensor Adaptive Processing (CAMSAP)}, pages 639--643. IEEE,
  2019.

\bibitem{larocca2021diagnosing}
Martin Larocca, Piotr Czarnik, Kunal Sharma, Gopikrishnan Muraleedharan,
  Patrick~J Coles, and M~Cerezo.
\newblock Diagnosing barren plateaus with tools from quantum optimal control.
\newblock {\em Quantum}, 6:824, 2022.

\bibitem{cerezo2021cost}
Marco Cerezo, Akira Sone, Tyler Volkoff, Lukasz Cincio, and Patrick~J Coles.
\newblock Cost function dependent barren plateaus in shallow parametrized
  quantum circuits.
\newblock {\em Nature communications}, 12(1):1--12, 2021.

\bibitem{grant2019initialization}
Edward Grant, Leonard Wossnig, Mateusz Ostaszewski, and Marcello Benedetti.
\newblock An initialization strategy for addressing barren plateaus in
  parametrized quantum circuits.
\newblock {\em Quantum}, 3:214, 2019.

\bibitem{belkin2018reconciling}
Mikhail Belkin, Daniel Hsu, Siyuan Ma, and Soumik Mandal.
\newblock Reconciling modern machine-learning practice and the classical
  bias--variance trade-off.
\newblock {\em Proceedings of the National Academy of Sciences},
  116(32):15849--15854, 2019.

\bibitem{ghorbani2019interpretation}
Amirata Ghorbani, Abubakar Abid, and James Zou.
\newblock Interpretation of neural networks is fragile.
\newblock In {\em Proceedings of the AAAI conference on artificial
  intelligence}, volume~33, pages 3681--3688, 2019.

\bibitem{chen2021equivalence}
Yilan Chen, Wei Huang, Lam Nguyen, and Tsui-Wei Weng.
\newblock On the equivalence between neural network and support vector machine.
\newblock {\em Advances in Neural Information Processing Systems},
  34:23478--23490, 2021.

\bibitem{el2010spectrum}
Noureddine El~Karoui.
\newblock The spectrum of kernel random matrices.
\newblock {\em The Annals of Statistics}, 38(1):1--50, 2010.

\bibitem{bergholm2018pennylane}
Ville Bergholm, Josh Izaac, Maria Schuld, Christian Gogolin, M~Sohaib Alam,
  Shahnawaz Ahmed, Juan~Miguel Arrazola, Carsten Blank, Alain Delgado, Soran
  Jahangiri, et~al.
\newblock Pennylane: Automatic differentiation of hybrid quantum-classical
  computations.
\newblock {\em arXiv preprint arXiv:1811.04968}, 2018.

\bibitem{jax2018github}
James Bradbury, Roy Frostig, Peter Hawkins, Matthew~James Johnson, Chris Leary,
  Dougal Maclaurin, George Necula, Adam Paszke, Jake Vander{P}las, Skye
  Wanderman-{M}ilne, and Qiao Zhang.
\newblock {JAX}: composable transformations of {P}ython+{N}um{P}y programs,
  2018.

\end{thebibliography}

% \appendix DOES NOT WORK IN IEEE
\appendix

\section{Theoretical and implementational details of the Path Kernel in the classical machine learning domain}\label{apx:path}

The Path Kernel was introduced in \cite{domingos2020every} as a means of replicating arbitrary gradient-descent based machine learning models in the form of kernel machines, under some weak assumptions. The Path Kernel is consequently of inherent interest in the theory of classical machine learning in that it grants a further layer of interpretability to models, including those, such as the neural networks, that often lacks this \cite{ghorbani2019interpretation}. In contrast, kernel machines permit a clear interpretation of prediction functions in terms of linear combinations of data in the training set as a consequence of the Representer Theorem. In particular, \cite[Theorem 1]{domingos2020every} indicates that the model $f(\vecx; \vecw) : \R^D \times \R^P \to \R$ (with $D$ the dimensionality of the data and $P$ the number of model parameters) can be rewritten:
\begin{equation}\label{eq:pk_appendix}
    f(\vecx; \bar{\vecw}) = \sum_{i=1}^m w_i(\vecx) \, K_\mathrm{path}(\vecx, \vecx_i; \bar\gamma) + w_0(\vecx).
\end{equation}
where 
\begin{align}
    K_\text{path} & : \R^D \times \R^D \times ([0,T]\to\R^P) \to \R^D \\
    K_\text{path}(\vecx, \vecx_i, \gamma) & = \int_{0}^{T} K_\text{tang}(\vecx, \vecx_i, \gamma(t)) \cdot \gamma'(t) dt
\end{align}
is the Path Kernel, a parametric kernel function (this parameterization has been rendered explicit in current formulation). In this case, $\bar\gamma : [0, T] \to \R^P$ is the parameter path as detailed in Section \ref{sec:quantumpathkernel} with a terminal parameter value $\bar{\gamma}(T) = \bar{\vecw}$. The Neural Tangent Kernel can also be expressed as a parametric kernel, 
\begin{align}
    K_\text{tang} & : \R^D \times \R^D \times \R^P \to \R \\
    K_\text{tang}(\vecx, \vecx_i; \vecw) & = \nabla_\vecw f(\vecx, \vecw) \cdot \nabla_\vecw f(\vecx_i, \vecw).
\end{align}
Equation \ref{eq:pk_appendix} holds under the proviso that $f$ is differentiable in $\vecw$,  and trained via Gradient Descent (GD) for the  given training dataset $\{(\vecx_i, y_i^*)\}_{i=1}^m \subseteq \R^D \times \R$ using the convex differentiable loss function $L(w) = \sum_{i=1}^M \ell(f(\vecx_i), y_i^*)$.

Equation~\ref{eq:pk_appendix} differs from a linear model due to the explicit dependency of the data $\vecx$ in the weights $w_i$, and it remains a matter of discussion as whether the path kernel in fact represents a more generalized model class  than   that of kernel machines (although it is clearly equivalent for infinitely small learning rates \cite{chen2021equivalence}). This debate need not concern us for the present purposes, where the intent is to obtain a class of models capable of representing the network gradient trajectory in a manner expressible on current quantum computers.   

As the Path Kernel is not widely deployed in practical machine learning, we detail here some of its properties. In \ref{apx:pk:mercer} we prove the Path Kernel is a Mercer Kernel. In \ref{apx:pk:domingo} we briefly comment on  the proof of \cite[Theorem 1]{domingos2020every}. In \ref{apx:pk:implementation} we demonstrate a numerical implementation of the Path Kernel.

\subsection{Path Kernel is a Mercer Kernel}\label{apx:pk:mercer}

Given any $\bar{\gamma}$, the function $\bar K_\text{path}(\vecx, \vecx') = K_\text{path}(\vecx, \vecx'; \bar\gamma)$ is a positive definite or Mercer kernel on $\R^D$. A Mercer kernel satisfies
\begin{equation}
    \sum_{i=1}^n \sum_{j=1}^n c_i c_j K(\vecx_i, \vecx_j) \ge 0
\end{equation}
for all sequence of elements $\vecx_1, ..., \vecx_n \in \R^D$ and constants $c_1, ..., c_n \in \R$. 

It is straightforward to demonstrate that such a condition is valid of the Path Kernel. Firstly, $\bar{K}_\text{tang}(\vecx, \vecx_i) = K_\text{tang}(\vecx, \vecx_i; \vecw)$ is a positive definite function for any $\vecw$ in consequence of  the positive definiteness of the Gram matrix of inner products in its parameter space $\R^P$. Secondly, since both the positive combination and the infinitesimal limit of combinations of positive definite kernels still satisfy the Mercer condition, then the preceding is immediately valid for the Path Kernel in both its discrete and continuous formulations.

\subsection{Comment on Theorem 1 in Domingo's work}\label{apx:pk:domingo}

In this section we  comment on \cite[Theorem 1]{domingos2020every} in order to highlight some of its limitations.
The dynamics of any predictor under training via gradient descent may be described by a first-order non-homogeneous differential equation:
\begin{equation}
    \frac{df(\vecx; \vecw)}{dt} = - \sum_{j=1}^P \frac{\partial f}{\partial w_j} \cdot \frac{\partial L}{\partial w_j}.
\end{equation}
where $f(\vecx; \vecw) : \R^D \times \R^P$ and $L$ is the convex differentiable loss function. 
We can describe these predictor dynamics over training in terms of the Tangent Kernel:
\begin{align}
    \label{th1:1}
    \frac{df(\vecx; \vecw(t))}{dt} 
    & = \sum_{j=1}^d \frac{\partial f(\vecx; \vecw)}{\partial w_j} \cdot \frac{d w_j}{dt} \\
    \label{th1:2}
    & = \sum_{j=1}^d \frac{\partial f(\vecx; \vecw)}{\partial w_j} \cdot \left( - \frac{\partial L(w(t))}{\partial w_j} \right) \\
    \label{th1:3}
    & = \sum_{j=1}^d \frac{\partial f(\vecx; \vecw)}{\partial w_j} \cdot \left( - \sum_{i=1}^m \frac{\partial \ell(y_i^*, f(x_i; w))}{\partial w_j} \right) \\
    \label{th1:4}
    & = \sum_{j=1}^d \frac{\partial f(\vecx; \vecw)}{\partial w_j} \cdot \left( - \sum_{i=1}^m \frac{\partial \ell(y_i^*, y_i)}{\partial y_i} \frac{\partial f(x_i; w)}{\partial w_j} \right) \\
    \label{th1:5}
    & = - \sum_{i=1}^m \frac{\partial \ell}{\partial y_i} \sum_{j=1}^d \frac{\partial f(\vecx; \vecw)}{\partial w_j} \frac{\partial f(\vecx_i; \vecw)}{\partial w_j} \\
    \label{th1:6}
    & = - \sum_{i=1}^m \frac{\partial \ell}{\partial y_i} \; \nabla_w f(\vecx; \vecw) \cdot \nabla_w f(\vecx_i; \vecw) \\
    \label{th1:7}
    & = - \sum_{i=1}^m \frac{\partial \ell}{\partial y_i} \; K_\text{tang}(\vecx, \vecx_i; \vecw)
\end{align}
In the limit $\epsilon \to 0$ we obtain:
\begin{multline}\label{eq:latest_pk}
    f(\vecx) = f(\vecx; \gamma(T))
    = f(\vecx; \gamma(0)) - \int_{0}^T \sum_{i=1}^m \frac{\partial \ell}{\partial y_i} \; K_\text{tang}(\vecx, \vecx_i; \gamma(t)) \; dt.
\end{multline}
Such a function cannot be straightforwardly represented as a linear model. However, by multiplying and dividing by the Path Kernel itself we obtain the following equation, at the cost of introducing a dependency of $\vecx$ in the model parameters:
\begin{align}
    f(\vecx; \gamma(T))
    & = f(\vecx; \gamma(0))
    + \sum_{i=1}^m 
    \left(- \frac{\int_{0}^T \frac{\partial \ell}{\partial y_i} \; K_\text{tang}(\vecx, \vecx_i; \gamma(t)) dt}{K_\text{path}(\vecx, \vecx_i; \gamma)}\right) 
    K_\text{path}(\vecx, \vecx_i; \gamma) \nonumber \\
    & = f(\vecx; \gamma(0)) + \sum_{i=1}^m 
    \alpha_i(\vecx) \, 
    K_\text{path}(\vecx, \vecx_i; \gamma).
\end{align}

Various works have suggested that imposing stronger assumptions on  training can remove the dependency of $\vecx$ in the model parameters. For example, the authors in \cite{chen2021equivalence} achieve this by imposing a requirement that the loss derivative is of  constant sign during training.

\subsection{Numerical calculation of the Path Kernel}\label{apx:pk:implementation}

We can calculate the value of the Path Kernel by approximating the integral with a direct sum
\begin{equation}
    K_\text{path}(\vecx, \vecx_i, \gamma) 
    = 
    \int_{0}^{T} K_\text{tang}(\vecx, \vecx_i, \gamma(t)) \cdot \gamma'(t) \, dt \approx 
    \sum_{t=0}^{T-1} K_\text{tang}(\vecx, \vecx_i, \gamma[t]) 
\end{equation}

\begin{figure*}[tbp]
    \centering
    \boxed{\includegraphics[width=0.5\textwidth]{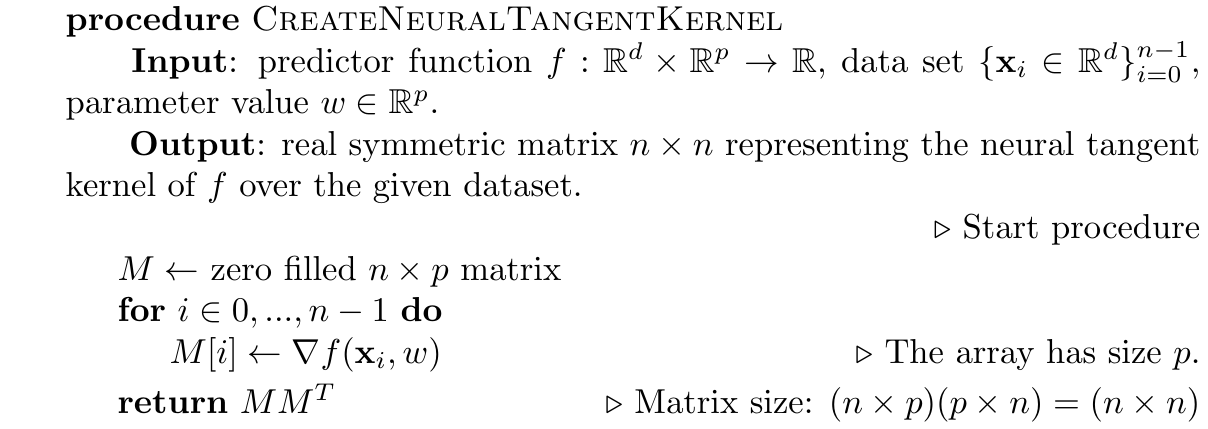}}
    \caption{Pseudo-code for the Neural Tangent Kernel formulation.}
    \label{algo:ntk}
\end{figure*}

The implementation details are reported in the following pseudo-code listings. In Figure~\ref{algo:ntk} we indicate how to calculate the Neural Tangent Kernel of the predictor $f$ once the parameter value $w$ is fixed. In particular, the gradient can be calculated with the finite difference method or, if the predictor is implemented with a Quantum Neural Network, with the parameter-shift rule.

\begin{figure*}[tbp]
    \centering
    \boxed{\includegraphics[width=0.5\textwidth]{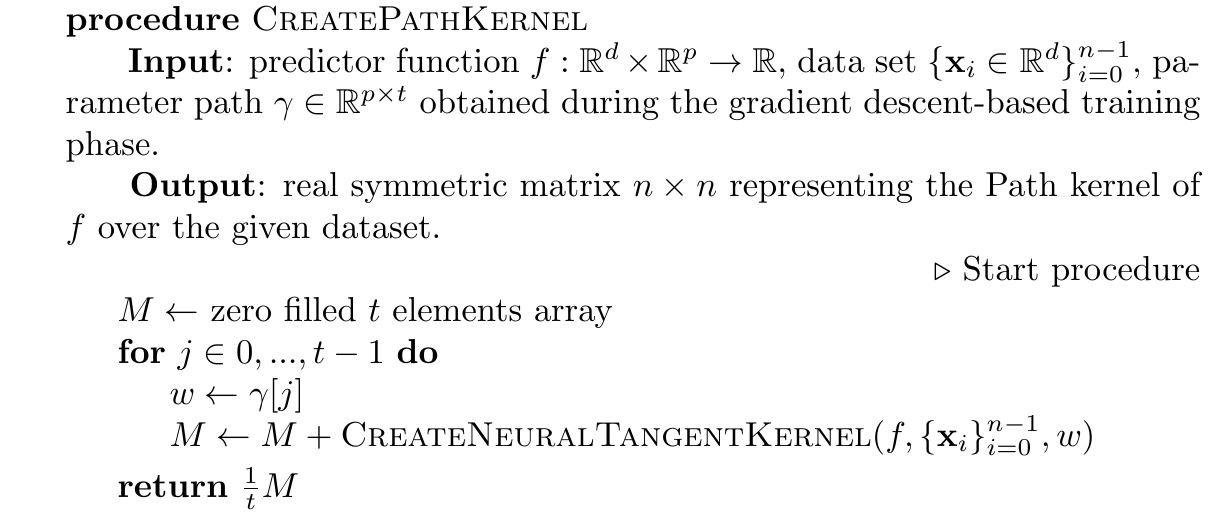}}
    \caption{Pseudo-code for the Path Kernel formulation.}
    \label{algo:pk}
\end{figure*}

\begin{figure*}[tbp]
    \centering
    \boxed{\includegraphics[width=0.5\textwidth]{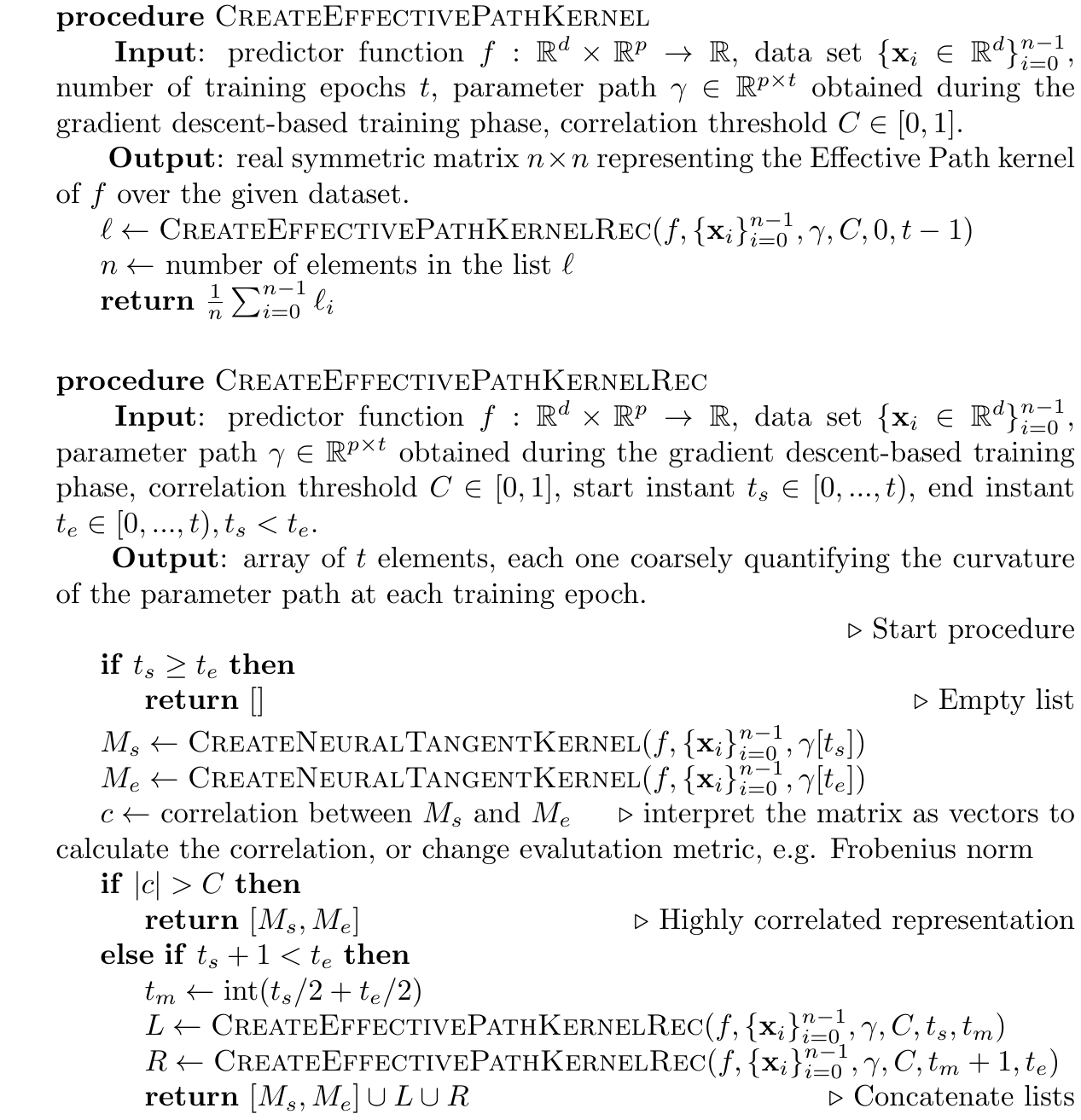}}
    \caption{Pseudo-code for the Effective Path Kernel formulation.}
    \label{algo:pk_effective}
\end{figure*}

The procedure for calculating the Path Kernel is shown in Figure \ref{algo:pk} and uses the Neural Tangent Kernel to calculate the individual contribution of each training epoch and thereafter calculates the average kernel matrix pointwise. 

In Section \ref{sec:quantumpathkernel:decorrelation} we discussed the potential significance of  decorrelated features; we here propose a numerical implementation of the \emph{Effective} Path Kernel. In contrast to the original Path Kernel, the Effective Path Kernel seeks to avoid to biasing due to multiple similar kernel contributions. This is especially important if the training has converged signficiantly earlier that  the last training epoch: any contribution after convergence has the same Neural Tangent Kernel and  will increase its relative weight as the number of epochs after convergence increases. Its formulation is given in Figure~\ref{algo:pk_effective}. Both the Path Kernel and Effective Path Kernel can be  straightforwardly implemented in parallel over multiple CPUs (or multiple QPUs) for the evaluation of $f$.

\section{Numerical evidence for the inability of random feature kernel techniques in solving the Gaussian XOR Mixture classification}\label{apx:refinetti}

In \cite{refinetti2021classifying} the authors demonstrate that a two-layer-depth neural network with only a small number of neurons can easily outperform kernel methods on the Gaussian Mixture classification problem, under the assumption that the number of training data points $n \to \infty$ is linearly proportional to the dimensionality of the data  $d \to \infty$.

We modify Refinetti's experiment for the current purposes to show the same result in a more straightforward way. 

We define the two-layer neural network as the function:
\begin{equation}\label{apx:eq:nn}
    f_\mathrm{nn}(\vecx; W_1, W_2, W_3, b_1, b_2, b_3) = W_3 \cdot \mathrm{relu}(W_2 \cdot \mathrm{relu}(W_1 \cdot \vecx + b_1) + b_2) + b_3
\end{equation}
parameterized by $W_1 \in \R^{h \times d}, W_2 \in \R^{h \times h}, W_3 \in \R^{1 \times h}, b_1, b2 \in \R^{h \times 1}, b_3 \in \R$, where $h$ is the number of hidden neurons per layer (the number of hidden neurons is here  fixed to $h = \lceil \sqrt{d} \rceil$). In our setting, we randomly initialize the weights $W_1, W_2, W_3$ by sampling the matrix element i.i.d. from a Gaussian of zero mean and unitary variance. The model is then trained using the gradient-descent-based algorithm ADAM for a maximum 1000 epochs with learning rate $0.001$ (the model is implemented in Python3 library scikit-learn, with the default configuration). 

We define a random feature kernel machine as:
\begin{equation}
    k_\mathrm{rf}(\vecx, \vecx') = \langle \phi(\vecx), \phi(\vecx') \rangle, \qquad \phi(\vecx) = \mathrm{relu}(W \cdot \vecx)
\end{equation}
with the activation weights parameterized by $W \in \R^{f \times d}, w_{i,j} \sim \mathcal{N}(0, 1)$, where $f$ has been chosen such that the number of parameters of the random feature kernel is greater that or equal to the number of parameters in the neural network, thus:
\begin{equation}
    f = \frac{(dh + hh + h) + (h + h + 1)}{d}.
\end{equation}
For $h = \lceil \sqrt{d} \rceil$ we can tightly upper bound $f$ with $f < \lceil \sqrt{d} \rceil + 5$. This kernel function is then fed to a SVM for classification (as  implemented in scikit-learn).

We randomly generate the dataset $\D_{d, d', \bar\epsilon, n}$ as detailed in Section \ref{sec:experiment:setup}. The experiment described below consists in comparing the performance of the neural network classifier with variations of the random feature kernel on the dataset $\D_{d, 3, \epsilon, 16d}$ for data point dimensionality $d = 4, 8, 12, 16, 20$ and noise $\epsilon = 0, 0.1, 0.2, ..., 1.9, 2.0$. We keep the number of non-zero features $d'=3$, meaning we are effectively classifying 3D Gaussian XOR mixtures, with the number of training vectors  of the dataset fixed to be $16 d$. The dataset is then randomly split $75\%$ in the training dataset and $25\%$ in the test set. For each dataset, we compare the performances of the oracle with the performances of the best of 10 randomly initialized neural networks and the best of 10 random feature kernels. For each dataset specification, we repeat this procedure $10$ times. 

\begin{figure*}[tbp]
    \centering
    % =============================================
    \begin{subfigure}[b]{0.3\textwidth}
    \includegraphics[width=\textwidth]{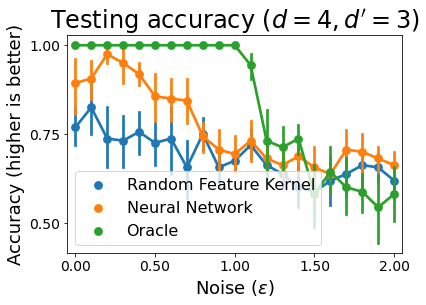}
    \caption{}
    \label{fig:ref:04}
    \end{subfigure}
    % =============================================
    \begin{subfigure}[b]{0.3\textwidth}
    \includegraphics[width=\textwidth]{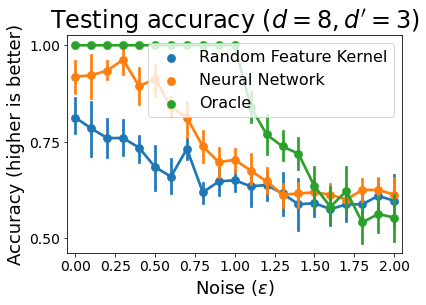}
    \caption{}
    \label{fig:ref:08}
    \end{subfigure}
    % =============================================
    \begin{subfigure}[b]{0.3\textwidth}
    \includegraphics[width=\textwidth]{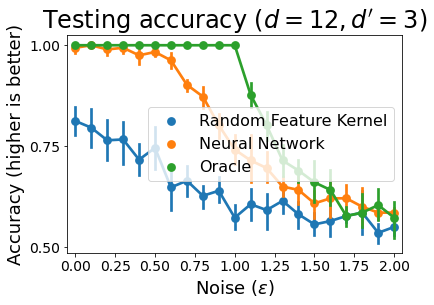}
    \caption{}
    \label{fig:ref:12}
    \end{subfigure}
    % =============================================
    \begin{subfigure}[b]{0.3\textwidth}
    \includegraphics[width=\textwidth]{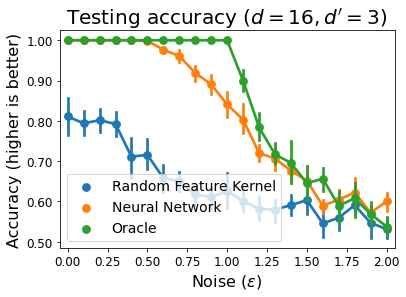}
    \caption{}
    \label{fig:ref:16}
    \end{subfigure}
    % =============================================
    \begin{subfigure}[b]{0.3\textwidth}
    \includegraphics[width=\textwidth]{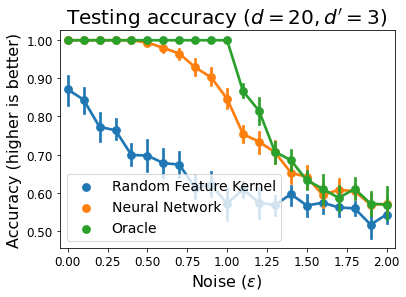}
    \caption{}
    \label{fig:ref:20}
    \end{subfigure}
    % =============================================
    \begin{subfigure}[b]{0.3\textwidth}
    \includegraphics[width=\textwidth]{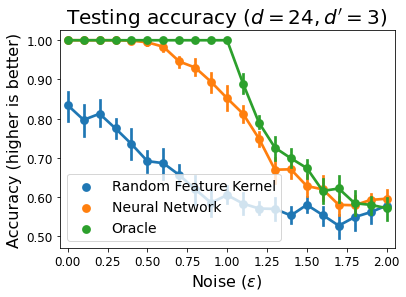}
    \caption{}
    \label{fig:ref:24}
    \end{subfigure}
    % =============================================
    \caption{Comparison of the performance of Random Feature Kernel and (2 layer) Neural Networks over the 3D Gaussian XOR Mixture problem with an increasing number of features set to zero. \ref{fig:ref:04}, \ref{fig:ref:08}, \ref{fig:ref:12}, \ref{fig:ref:16}, \ref{fig:ref:20}, \ref{fig:ref:24} have respectively 4, 8, 12, 16, 20, 24 feature per point, the first three being the only non-zero ones.}
    \label{fig:refinetti}
\end{figure*}

\begin{figure*}[tbp]
    \centering
    % =============================================
    \begin{subfigure}[b]{0.3\textwidth}
    \includegraphics[width=\textwidth]{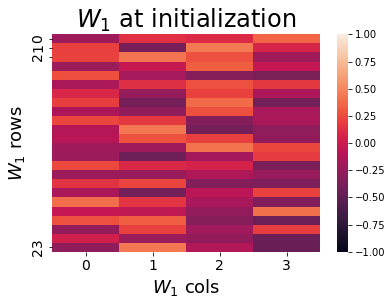}
    \caption{}
    \label{fig:nn:000}
    \end{subfigure}
    % =============================================
    \begin{subfigure}[b]{0.3\textwidth}
    \includegraphics[width=\textwidth]{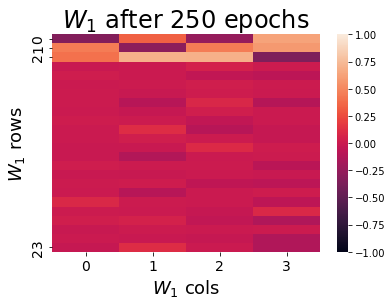}
    \caption{}
    \label{fig:nn:250}
    \end{subfigure}
    % =============================================
    \begin{subfigure}[b]{0.3\textwidth}
    \includegraphics[width=\textwidth]{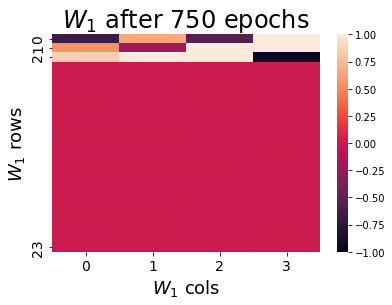}
    \caption{}
    \label{fig:nn:750}
    \end{subfigure}
    % =============================================
    \caption{Values of the $W_1$ matrix for individual neural networks of the form of Equation~\ref{apx:eq:nn} during training on  the Gaussian XOR Mixture datasets $\D_{24, 3, 0.8, 384}$: \ref{fig:nn:000}, \ref{fig:nn:250}, \ref{fig:nn:750} represent the coefficients at initialization, after 250 training epochs and after 750 training epochs of training with ADAM at a learning rate $0.001$.}
    \label{fig:nn_weights}
\end{figure*}

In Figure~\ref{fig:refinetti} we set out   the results of the above described experiments. It may be observed that  Neural Networks  outperform the kernel approach in each case, with the differential in accuracy increasing with the number of zero-valued features. Refinetti et al. \cite{refinetti2021classifying} suggest that this difference in performance is accounted for by the fact that random feature kernels in high dimension behave as linear transformations \cite{el2010spectrum}.

We have here suggested a complementary interpretation of the results of such experiments. We have shown that the difference of performance between the two models is not uniquely determined by the failure of kernel methods \emph{per se}. In fact, it is determined also by the feature learning capabilities of neural networks; inspecting the evolution of the $W_1$ parameters during the training of a neural network reveals that elements in $W_1$ related to the zero-features do indeed go to zero (Figure~\ref{fig:nn_weights}). This results in having all of the hidden neurons (whose number is proportional to $\sqrt{d}$ and thus increasing with the number of features) working adaptively to classify the three discriminatively informative components or features, thereby improving overall performance in contrast to the random feature kernel approach, for which adding feature (and parameters) drastically decreases  performance (which is to say the path model outperforms the random feature kernel in this problem  by being able to discharge junk features candidates,  thus performing feature learning).

\section{Data, Code, and Simulation details}\label{appendix:simulation}

Both the code to reproduce the indicated experiments and also the relevant data are freely available at  \texttt{https://github.com/incud/QuantumPathKernel}. The code is released open-source. 

The indicated experiments have been simulated on two devices:
\begin{itemize}
    \item one Dell Latitude 5510 having: Intel Core i7-10610U CPU with 4 physical cores, 16GB RAM, without CUDA-enabled GPUs;
    \item one cluster node having: Intel Xeon Silver 4216 CPU with 64 physical cores, 180GB RAM, with 4 x CUDA-enabled GPUs NVidia Tesla V100S 32GB VRAM.
\end{itemize}
The software runs on Ubuntu 20.04 LTS and uses Python v3.9.11, PiP packet-manager v22.0.4 along with the other libraries listed in \texttt{requirements.txt} file in the root of the attached repository. Installation and simulation instructions are documented in the \texttt{README.md} file in the root of the repository. Our code is based upon freely available, open-source frameworks only.

The framework used to define and simulate the quantum circuit is PennyLane \cite{bergholm2018pennylane}. The simulations have been accelerated using the JAX library \cite{jax2018github}. (JAX might require installation from source code if used on operating systems different from Ubuntu). Alternatively, the source code can be set such that PennyLane does not require this library. (However, in this case, the circuit simulation might be substantially slower and would not benefit the full potential of multicore CPUs and GPUs). These experiments have not been run on quantum hardware. 

The input and output of each experiment are contained in  different subfolders within the root directory. They contain the specifications needed to generate the training and testing datasets, the datasets themselves, the trace of the parameters during the training for any model, and the Quantum NTK and Quantum Path Kernel Gram matrices for each model (which may be used to create a pre-trained model), and also the resulting plots. The \texttt{README.md} explains in detail the commands needed to reproduce our results. 

The simulations for all experiments have taken approximately 600 hours across both machines used.

\end{document}